\documentclass[]{aa}

\usepackage{graphicx}
\usepackage{natbib}

\usepackage{amsmath,amssymb}
\usepackage{fixltx2e}
\usepackage{hyperref}
\usepackage{txfonts}
\usepackage[normalem]{ulem}

\begin{document}

\title{Detecting stellar-wind bubbles through infrared arcs in \ion{H}{ii} regions}
\author{
  Jonathan~Mackey\inst{1,2} \and
  Thomas J.~Haworth\inst{3} \and
  Vasilii V.~Gvaramadze\inst{4,5,6} \and
  Shazrene Mohamed\inst{7} \and
  Norbert Langer\inst{2} \and
  Tim J.~Harries\inst{8}
        }
\institute{
I.\ Physikalisches Institut, Universit\"at zu K\"oln, Z\"ulpicher Stra\ss{}e 77, 50937 K\"oln, Germany \\ \email{mackey@ph1.uni-koeln.de}
  \and  
  Argelander-Institut f\"ur Astronomie, Auf dem H\"ugel 71, 53121 Bonn, Germany
  \and
  Institute of Astronomy, Madingley Road, Cambridge, CB3 0HA, UK
  \and
  Sternberg Astronomical Institute, Lomonosov Moscow State University, Universitetskij Pr.~13, Moscow 119992, Russia
  \and
  Space Research Institute, Russian Academy of Sciences, Profsoyuznaya 84/32, 117997 Moscow, Russia
  \and
  Isaac Newton Institute of Chile, Moscow Branch, Universitetskij Pr.\ 13, Moscow 119992, Russia
  \and
  South African Astronomical Observatory, P.O.\ box 9, 7935 Observatory, South Africa
  \and
  Department of Physics and Astronomy, University of Exeter, Stocker Road, Exeter, EX4 4QL, UK
}

\date{Submitted: 2015.10.15 / Revised 2015.12.10 / Accepted 2015.12.20}

\abstract{
  Mid-infrared arcs of dust emission are often seen near ionizing stars within  \ion{H}{ii} regions.
  A possible explanations for these arcs is that they could show the outer edges of asymmetric stellar wind bubbles.
  We use two-dimensional, radiation-hydrodynamics simulations of wind bubbles within \ion{H}{ii} regions around individual stars to predict the infrared emission properties of the dust within the \ion{H}{ii} region.
  We assume that dust and gas are dynamically well-coupled and that dust properties (composition, size distribution) are the same in the \ion{H}{ii} region as outside it, and that the wind bubble contains no dust.
  We post-process the simulations to make synthetic intensity maps at infrared wavebands using the \textsc{torus} code.
  We find that the outer edge of a wind bubble emits brightly at 24 $\mu$m through starlight absorbed by dust grains and re-radiated thermally in the infrared.
  This produces a bright arc of emission for slowly moving stars that have asymmetric wind bubbles, even for cases where there is no bow shock or any corresponding feature in tracers of gas emission.
  The 24 $\mu$m intensity decreases exponentially from the arc with increasing distance from the star because the dust temperature decreases with distance.
  The size distribution and composition of the dust grains has quantitative but not qualitative effects on our results.
  Despite the simplifications of our model, we find good qualitative agreement with observations of the \ion{H}{ii} region RCW\,120, and can provide physical explanations for any quantitative differences.
  Our model produces an infrared arc with the same shape and size as the arc around CD\,$-$38$\degr$11636 in RCW\,120, and with comparable brightness.
  This suggests that infrared arcs around O stars in \ion{H}{ii} regions may be revealing the extent of stellar wind bubbles, although we have not excluded other explanations.
}

\keywords{
  Hydrodynamics -
  radiative transfer - 
  methods: numerical -
  H~\textsc{ii} regions -
  ISM: bubbles -
  Stars: winds, outflows -
  individual objects: \object{RCW\,120} -
  individual objects: \object{[CPA2006] N49}
}
\authorrunning{Mackey et al.}
\titlerunning{Infrared arcs as tracers of stellar wind bubbles}
\maketitle

\section{Introduction}
\label{sec:intro}

Stars with mass $M\gtrsim20\,\mathrm{M}_\sun$ (O stars) emit copious extreme-ultraviolet (EUV) photons capable of ionizing hydrogen when on the hydrogen-burning main sequence and also have line-driven stellar winds with terminal velocities $v_\infty\gtrsim1000\,\mathrm{km\,s}^{-1}$ \citep{SnoMor76}, with important consequences for their surroundings \citep{Dal15}.
Their EUV photons create overpressurised photoionized \ion{H}{ii} regions that expand and drive a dense shocked shell into the interstellar medium (ISM).
The stellar winds create an expanding cavity of shocked wind material that is very hot ($10^6-10^8\,$K) and rarefied, and which emits in X-rays \citep{Ave72, DysdeV72, CasMcCWea75, WeaMcCCasEA77}.
If this stellar wind bubble (SWB) expands supersonically within the surrounding \ion{H}{ii} region it will drive a shocked shell \citep{Ave72}, but this typically only occurs for young wind bubbles in dense molecular clouds where stars are born
\citep[$<10^4-10^5$ years;][]{DysdeV72}.
The SWB remains smaller than the \ion{H}{ii} region for much or all of the main sequence for all but the most massive O stars \citep{WeaMcCCasEA77, VanLanGar05, FreHenYor06}.
The \ion{H}{ii} region and wind bubble can become distorted if the star is moving \citep{WeaMcCCasEA77, RagNorCanEA97, MeyMacLanEA14}, and if the star moves supersonically through the \ion{H}{ii} region then the wind bubble drives a bow shock in the direction of motion \citep{BarKraKul71, WeaMcCCasEA77, ArtHoa06, ZhuZhuLiEA15}.
In \citet[][hereafter Paper I]{MacGvaMohEA15}, we showed that even slowly moving stars (space velocity, $v_\star=4\,\mathrm{km\,s}^{-1}$) produce very asymmetric wind bubbles in dense regions.

Direct detection of SWBs around main sequence O stars is difficult because they are filled with low-density, hot gas that typically has very low emission measure.
X-ray emission has been detected from SWBs around evolved, single, Wolf-Rayet stars \citep{Boc88, WriWenWis94, ChuGueGruEA03, ToaGueChuEA12, ToaGueGruEA14, Zhe14, ToaGueChuEA15}, because these stars have winds that are much denser than O stars \citep{Cro07}, and because the bubbles are dynamically quite young.
It has also been detected from high-mass-star forming regions containing multiple O stars \citep{TowFeiMonEA03, TowBroChuEA11, TowBroGarEA14}, where many wind bubbles have merged into a superbubble and supernovae may have occurred \citep{McCKaf87}.
If predictions for the X-ray luminosity of SWBs around single O stars are correct \citep[][Paper I]{Osk05, FreHenYor06, ToaArt11} then current X-ray telescopes will not detect their diffuse emission, although the next generation of telescopes may have sufficient sensitivity.
It is therefore useful to predict the morphology and spectral energy distribution of such emission \citep{Art12, RogPit14, KraDieBohEA14}, but it is important to also consider alternative, indirect, methods for detecting SWBs.

Indirect detection of SWBs is achieved by searching for ``holes'' in ISM emission around O stars, because the SWB is sufficiently low-density that it is effectively invisible.
Detection is easier if the bubble expands supersonically because the surrounding ISM is shocked, thus compressed, heated, and emitting more brightly for most tracers \citep[e.g.,][]{MeyMacLanEA14}, but this only occurs for supersonically moving stars or for very young SWBs that are expanding rapidly.
Detection is also easier if the surrounding ISM is dense, because emission/absorption features stand out more clearly against background and foreground fluctuations.

The most common tracers of photoionized gas are recombination lines (e.g.,~H$\alpha$, H$\beta$), collisionally excited forbidden lines of C, N, O, and S ions, and radio bremsstrahlung \citep[e.g.,][]{Spi78, WatPovChuEA08}.
Some \ion{H}{ii} regions do show a central decrement in emission near the ionizing star \citep[e.g.,~N49;][]{WatPovChuEA08} in these tracers, but others do not \citep[e.g.,~RCW\,120;][]{OchVerCoxEA14, TorHasHatEA15}.
This can be understood if some SWBs fill only a small fraction of the \ion{H}{ii} region along the line of sight (so that the fractional decrement is small), but we then need more sensitive tracers to detect these SWBs.

IR observations of thermal dust emission is an attractive possibility because winds from hot stars are dust-free, whereas \ion{H}{ii} regions contain a normal ISM dust distribution, with the exception of PAHs that are destroyed by ionizing radiation \citep{PovStoChuEA07, PavKirWie13}.
O stars are luminous radiation sources, and so even though only a small fraction of the stellar radiation is absorbed by dust, the re-emitted radiation is often more luminous than that of all gas-cooling radiation \citep{MeyMacLanEA14}.
This explains why infrared surveys \citep{VanNorDga95, GvaKroPfl10, GvaPflKro11, GvaKniKroEA11, PerBenBroEA12} have been much more successful at finding bow shocks than optical searches \citep{BroBow05}.
It may also indicate that the outer boundary of a SWB is more easily detected in IR than in optical observations.

Dust grains in radiative equilibrium are warmest close to the O star and their temperature decreases with distance.
At typical grain temperatures \citep[20-90 K;][]{PalUmaVenEA12} the 8 and 24 $\mu$m IR wavebands are in the Wien tail of the dust emission spectrum, and so emissivity is exponentially sensitive to temperature.
If there is an O star within a dust-free SWB surrounded by a constant-density ISM, we naively expect that the 24 $\mu$m emission from the dust will then peak at the SWB boundary and decrease exponentially with distance from the O star until the \ion{H}{ii} region boundary is reached.
For an asymmetric wind bubble, this will result in a bright arc of emission, with an O star near the focus of the arc.

A huge number of IR arcs and shells have been discovered by \textit{Spitzer} \citep[e.g.,][]{ChuPovAllEA06, DehSchAndEA10, GvaKniFab10, WacMauvDykEA10, MizKraFlaEA10, SimPovKenEA12, KenSimBreEA12}, a wealth of data showing that many bubbles have IR arcs within larger \ion{H}{ii} region shells.
\citet{WeaMcCCasEA77} showed that for most main sequence stars, it takes more than 1\,Myr before the SWB can sweep up enough mass to trap the \ion{H}{ii} region, so these new data require theoretical models including both SWBs and \ion{H}{ii} regions to be understood.
Examples of mid-IR arcs within \ion{H}{ii} regions include:
around HD 64315 in NGC 2467 \citep{SniHesDesEA09},
around \object{CD\,$-$38$\degr$11636} in RCW\,120 \citep[][Paper~I]{OchVerCoxEA14},
an arc within the \ion{H}{ii} region RCW\,82 \citep{OchVerCoxEA14},
around $\sigma$ Ori within IC\,434 \citep{OchCoxKriEA14},
around $\lambda$ Ori \citep{OchBroBalEA15},
an almost-complete ring in N49 \citep{WatPovChuEA08},
an arc within G31.165-00.127 \citep{DehSchAndEA10},
and various arcs in the sample of \citet{PalUmaVenEA12}.
We have excluded from this list arcs around runaway stars that are definitely bow shocks \citep[e.g.,][]{GvaLanMac12}.

So far the emission from these arcs has not been modelled using multidimensional RHD simulations.
They have been interpreted in the context of
(i) dust waves where gas and dust are dynamically decoupled by radiation pressure \citep[we refer to this as the \textit{decoupling model};][]{VanMcC88, OchVerCoxEA14},
(ii) a \textit{displacement model}, where a SWB evacuates dust from a region near the O star \citep{WatPovChuEA08, PavKirWie13}, and
(iii) a \textit{mixing model}, where dense, dusty clumps of ISM are photoevaporated, heated, and mixed with the shocked stellar wind \citep{McKVanLaz84}.
In this paper we investigate the displacement model as the origin of IR arcs within \ion{H}{ii} regions, to test whether they could be revealing the edges of asymmetric SWBs.
We make the assumptions that gas and dust are dynamically coupled and that dust properties (composition, size distribution) are unaffected by proximity to the ionizing O star.
IR emission maps for these assumptions have only been made for spherically symmetric \ion{H}{ii} regions that do not include a stellar wind \citep{PavKirWie13}.
We take results from two-dimensional (2D), radiation-hydrodynamics (RHD) simulations of wind bubbles within \ion{H}{ii} regions (Paper I) and post-process them to make synthetic maps of thermal dust emission.
We then quantitatively compare our predictions to \textit{Spitzer} and \textit{Herschel} observations of the \ion{H}{ii} region RCW\,120, to see if the observed IR arc is compatible with the edge of a SWB.

Sect.~\ref{sec:sims} introduces the RHD simulations from Paper I and some additional calculations that we have made for this work.
In Sect.~\ref{sec:dust} we describe the method for modelling dust emission, including the physical model and the radiative transfer methods.
The synthetic emission maps are presented and compared to observations in Sect.~\ref{sec:results}.
We discuss our findings in the context of previous work in Sect.~\ref{sec:discussion} and present our conclusions in Sect.~\ref{sec:conclusions}.

\section{Simulations} \label{sec:sims}

\subsection{Methods and initial conditions}

In Paper I we presented radiation-hydrodynamic simulations modelling the simultaneous evolution of an \ion{H}{ii} region and wind bubble around a slowly moving O star.
Here we take simulation `V04' from Paper I for a star moving with a space velocity $v_\star=4\ \mathrm{km\,s}^{-1}$ through the ISM, rename it WV04, and supplement it with another simulation denoted HV04, identical to WV04 except that there is no wind, only an \ion{H}{ii} region.
To these we add two new simulations, HV00 and WV00, the same except for a static star ($v_\star=0$) and that they are run with spherical symmetry (i.e.\ one-dimensional calculations).
All four simulations have the same spatial resolution (256 cells per parsec), and we have also run lower resolution simulations with 128 cells per parsec to check that the simulation properties have converged.
The ISM has a uniform H number density of $n_\mathrm{H}=3000$ cm$^{-3}$ (the mean number density of the ISM around RCW\,120, \citealt{ZavPomDehEA07}).
The properties of the simulations are described in Table~\ref{tab:sims}.
All simulations are run with the radiation-magnetohydrodynamics code \textsc{pion} \citep{MacLim10, Mac12}.
The Euler equations are solved on a uniform grid with a finite-volume, shock-capturing scheme that is accurate to second order in time and space, and which does not require operator splitting for source terms \citep{Fal91, FalKomJoa98}.
The H ionization fraction, $x$ is advected as a passive tracer, and stellar wind material is distinguished from ISM by a second passive tracer.

\begin{table}
  \centering
  \caption{
    Simulations used for post-processing.
    Simulations without a stellar wind have an ID prefixed with `H' (for \ion{H}{ii} region), and simulations with both stellar wind and ionizing radiation have an ID prefixed with `W' (for wind).
    All simulations use the same ionizing radiation source; see text for details.
    The simulations with a stellar wind also both use the same wind, described in the text.
    Dim.\ is the number of dimensions used for the simulation (1D uses spherical symmetry, 2D uses axisymmetry);
    $v_\star$ is the star's space velocity in $\mathrm{km}\,\mathrm{s}^{-1}$;
    $N_\mathrm{zones}$ shows the number of grid zones in the simulation (radial for 1D, and $N_\mathrm{z}\times N_\mathrm{R}$ in 2D, being the number of grid zones in the $\mathbf{\hat{z}}$ and $\mathbf{\hat{R}}$ directions, respectively);
    the last column shows the simulation domain size in pc.
    Simulation WV04 was labelled `V04' in Paper I; the other simulations were not presented in Paper I.
  }
  \begin{tabular}{ l c c c c c}
    ID & Dim. & wind?  & $v_*$ & $N_\mathrm{zones}$ & Domain (pc) \\
    \hline
    HV00 & 1 & no  & 0 & $512$  & $2.0$  \\
    WV00 & 1 & yes & 0 & $512$  & $2.0$  \\
    HV04 & 2 & no  & 4 & $1280\times512$  & $5.0\times2.0$  \\
    WV04 & 2 & yes & 4 & $1280\times512$  & $5.0\times2.0$  \\
  \end{tabular}
  \label{tab:sims}
\end{table}

We consider the star as a point source of ionizing (EUV) and non-ionizing (FUV) photons and of a spherically symmetric wind, fixed at the origin.
A short-characteristics raytracer \citep{RagMelArtEA99} is used to calculate photon attenuation, and a high-order implicit integrator \citep[\textsc{cvode};][]{CohHin96} is used to integrate the radiative heating/cooling together with the non-equilibrium ionization of H.
Photon conservation is ensured by using an appropriate formulation of the photoionization rate \citep{AbeNorMad99, MelIliAlvEA06} together with timestepping restrictions described for Algorithm 3 in \citet{Mac12}.
We use the on-the-spot approximation (local absorption of scattered ionizing photons) and do not include dust absorption of EUV photons.
For FUV photons, we assume dust is the only source of opacity.
Gas heating and cooling are as described in Paper I, partly non-equilibrium and partly based on cooling curves and approximate functions.

The star is taken to emit a blackbody spectrum with effective temperature $T_\mathrm{eff}=37\,500$ K, and has an EUV photon luminosity $Q_{0}= 3\times10^{48}$ s$^{-1}$ and FUV photon luminosity $Q_{\mathrm{FUV}}=7.5\times10^{48}$ s$^{-1}$.
Its wind is characterised by a mass-loss rate of $\dot{M}=1.55\times10^{-7} \,\mathrm{M}_\sun\,\mathrm{yr}^{-1}$ and terminal velocity $v_\infty=2000\,\mathrm{km}\,\mathrm{s}^{-1}$.
These values are appropriate for a Galactic O star with mass $M\approx30\ \mathrm{M}_\sun$ \citep{MarSchHil05};
the radiative properties are those determined for CD\,$-$38$\degr$11636 by \citet{MarPomDehEA10}.
The wind properties have been taken from \citet{VinDeKLam01}, and we will show that they produce a 24 $\mu$m arc that is the same size as the arc around CD\,$-$38$\degr$11636.

\begin{figure*}[ht]
\includegraphics[width=0.49\textwidth]{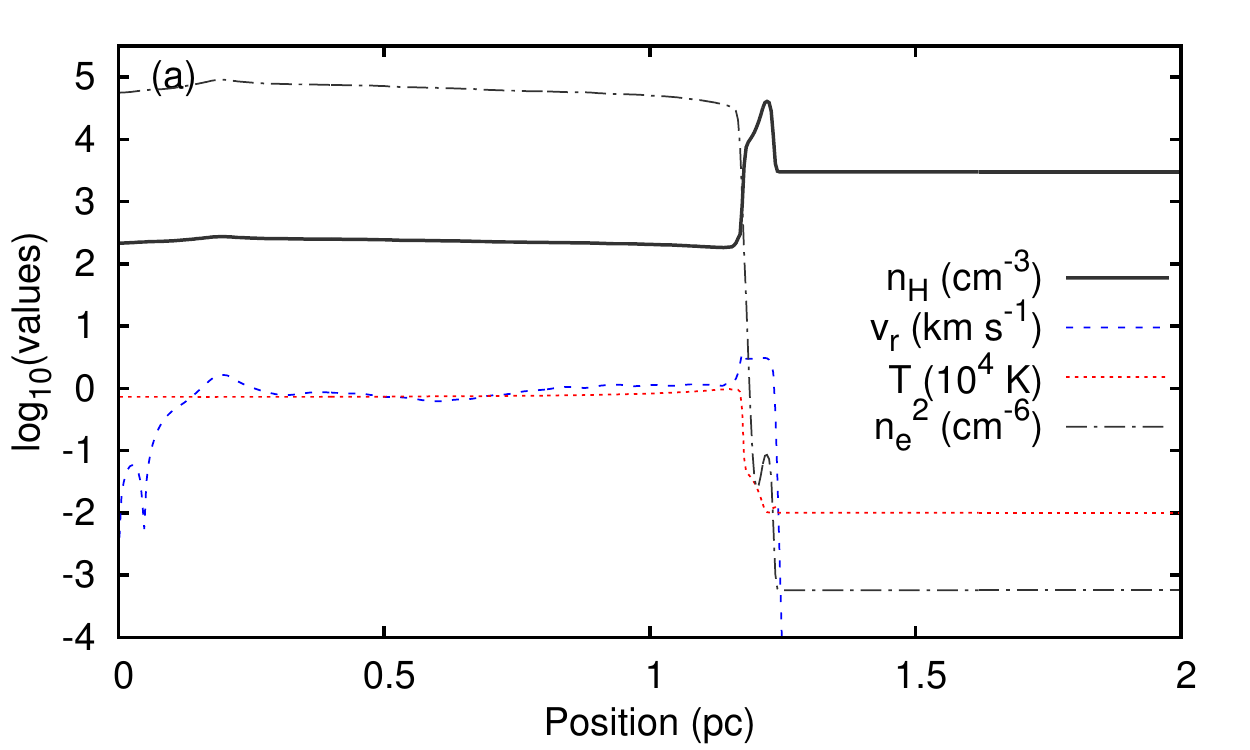}
\includegraphics[width=0.49\textwidth]{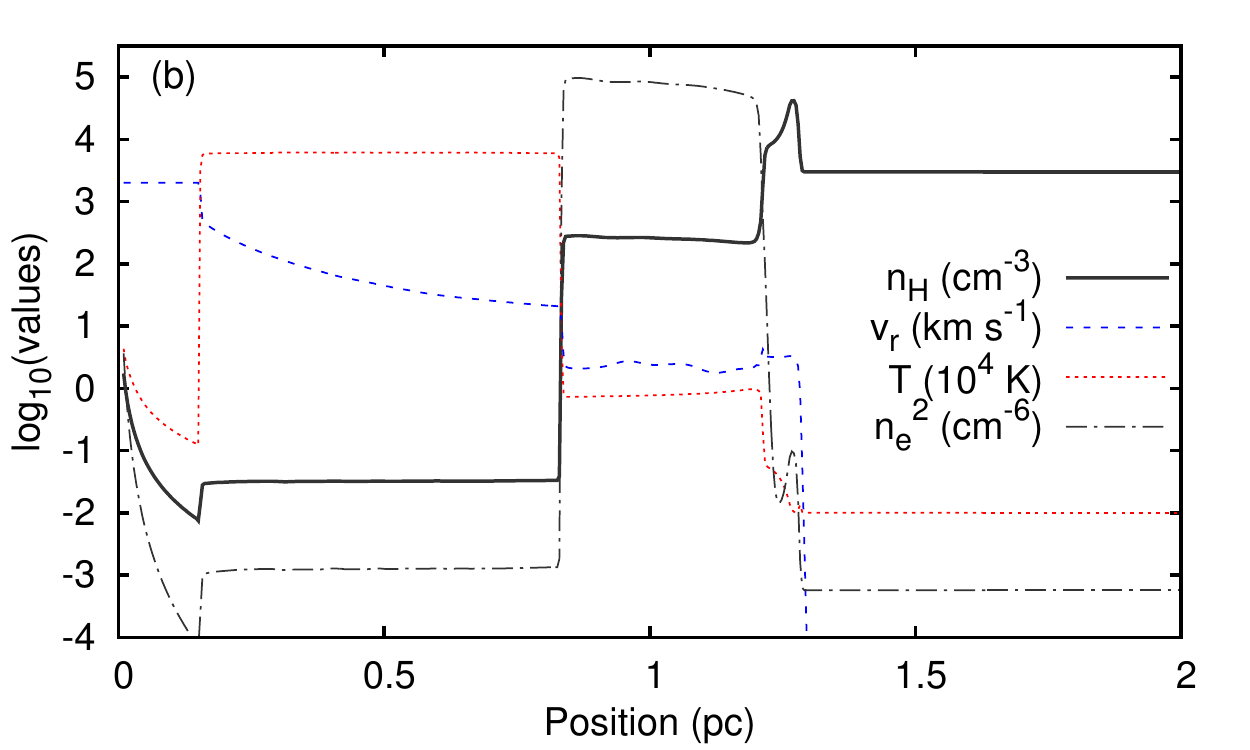}
\caption{
  Simulation HV00 (a) and WV00 (b) showing log of: $n_{\mathrm{H}}$, expansion velocity $v_\mathrm{r}$, temperature, T, and square of the electron number density, $n_\mathrm{e}^2$, with units as indicated in the legend.
  Both simulations are plotted at time $t=0.2$ Myr.
}
\label{fig:V00slice}
\end{figure*}

\begin{figure*}
\includegraphics[width=0.49\textwidth]{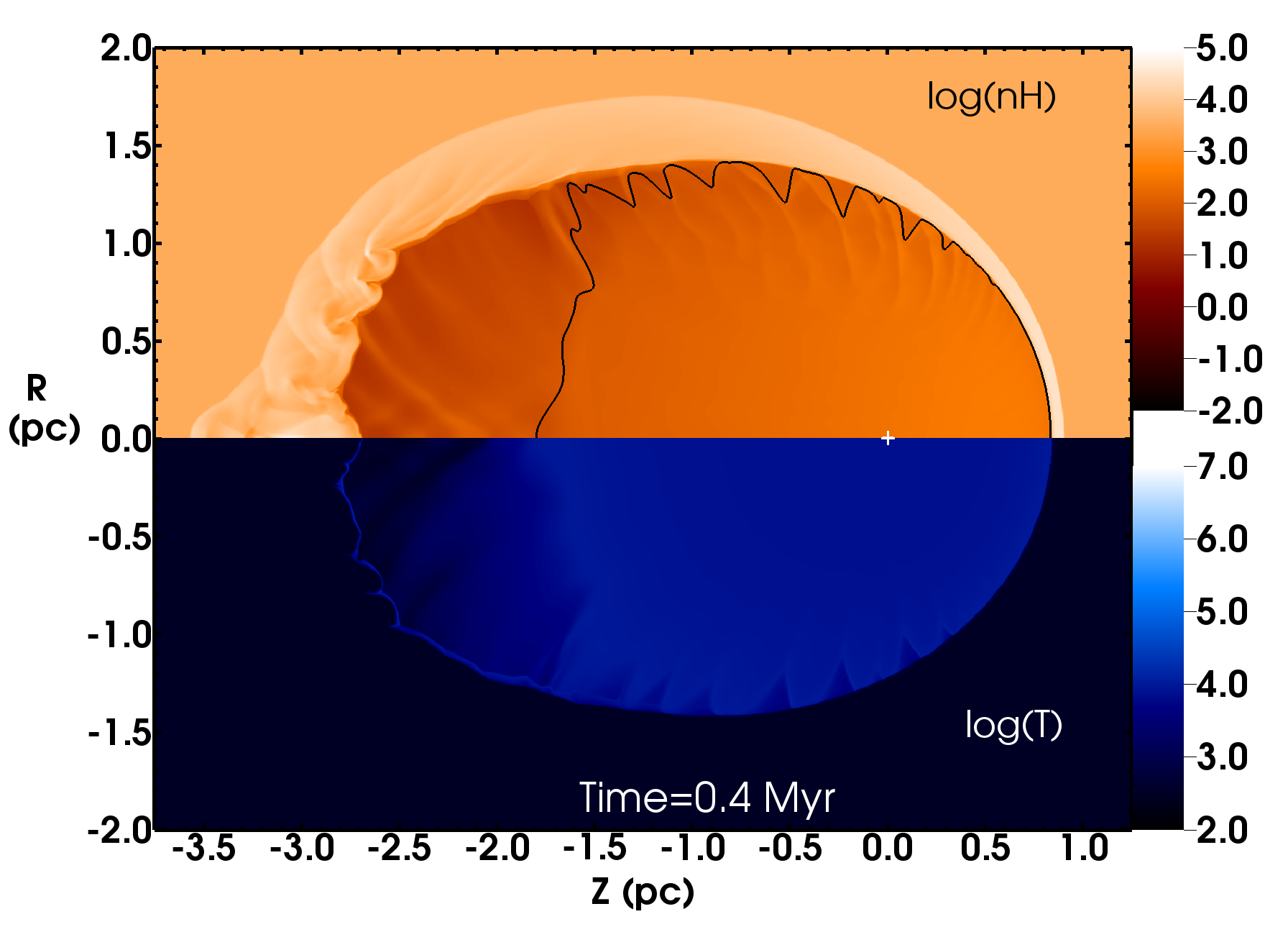}
\includegraphics[width=0.49\textwidth]{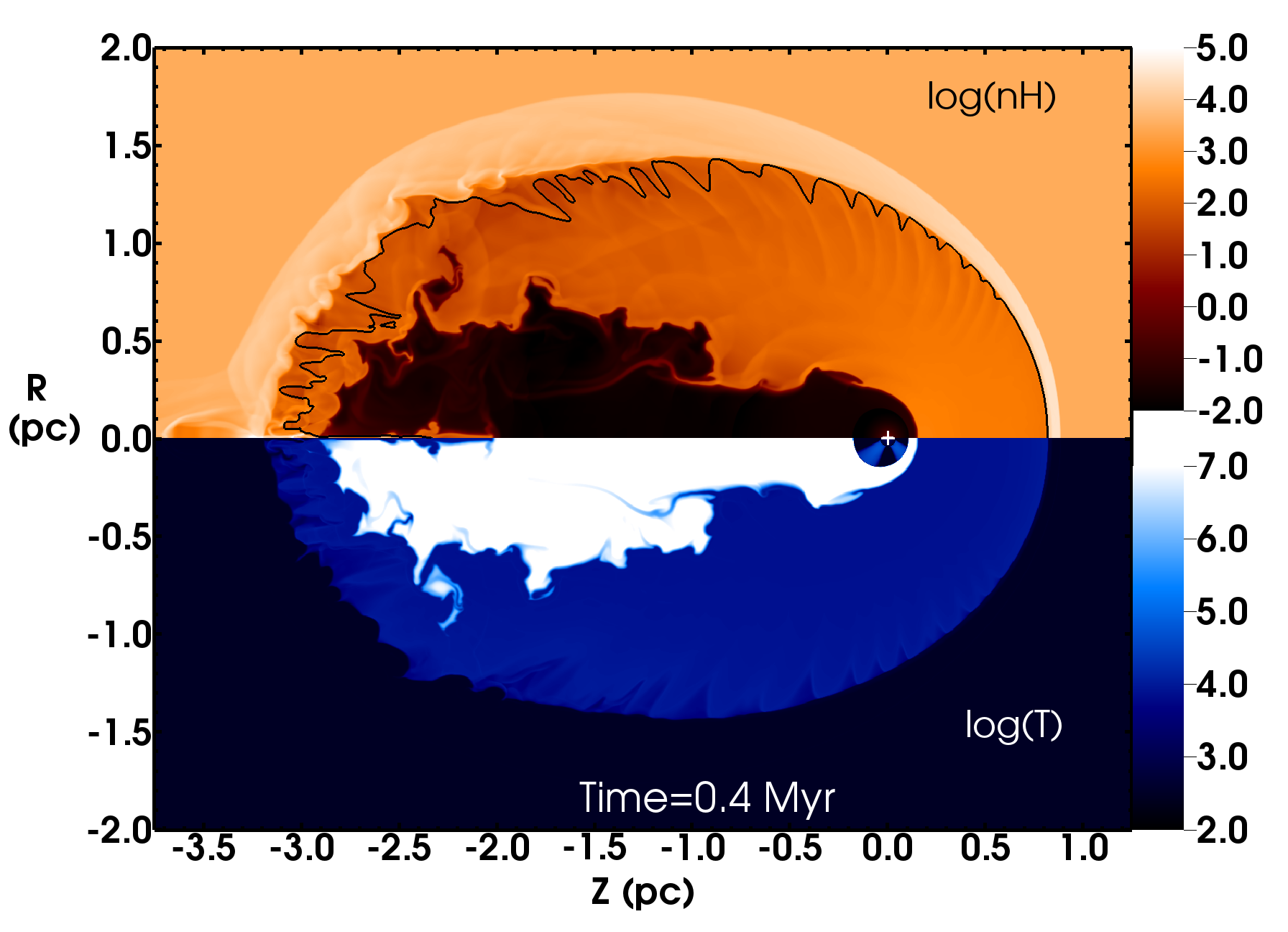}
\caption{
  Snapshots of the image plane for simulation HV04 (left) and WV04 (right) showing log of gas number density (upper half-plane) and temperature (lower half-plane).
  The units are $\log(n_{\mathrm{H}}{}/\mathrm{cm}^{-3})$ and $\log(T/\mathrm{K})$, respectively.
  A contour showing H ionization fraction, $x=0.5$, is also plotted in black in the upper half-plane.
  The symmetry axis is horizontal, labelled $z$; the star is at the origin indicated by the white cross, and the ISM is flowing past from right to left at $4\ \mathrm{km\,s}^{-1}$.
  Both simulations are plotted at time $t=0.4$ Myr.
}
\label{fig:V04slice}
\end{figure*}

\subsection{Simulation snapshots}

Snapshots from simulations HV00 and WV00 are shown in Fig.~(\ref{fig:V00slice}) after 0.2\,Myr of evolution, and from simulations HV04 and WV04 in Fig.~(\ref{fig:V04slice}) after 0.4\,Myr of evolution.
All of the synthetic emission maps in this paper use these snapshots.
At this time the simulations HV04 and WV04 are close to reaching a stationary state in the upstream ($z>0$) direction.
In this paper we are mainly concerned with one-sided arcs of emission, which are formed by asymmetric pressure arising from relative motion between the star and the ISM.
For this case, especially for the case of dense medium, the hydrodynamics reaches a steady state in a time short compared to the evolutionary timescales of the O star (see paper I).
In contrast, HV00 and WV00 will not reach a steady state until after the \ion{H}{ii} region internal pressure equals the external pressure \citep{RagCanRodEA12, BisHawWilEA15}.
This can take longer than the main sequence lifetime of an O star \citep[e.g.][]{VanLanGar05, FreHenYor06}.

For HV00 and WV00 the simulations are spherically symmetric so we plot various quantities as a function of radius from the star.
The \ion{H}{ii} region properties are not substantially affected by the stellar wind, except that there is a cavity of low-density gas around the star.
The location of the ionization front and the dense shell driven by \ion{H}{ii} region expansion are almost unaffected, and the density of gas within the \ion{H}{ii} region is very similar.
The volume of the SWB in WV00 is somewhat dependent on resolution, because it is very sensitive to numerical mixing of the hot bubble with the cooler and denser ISM \citep{MacMcCNor89, VanKep11}.
Note that the emission measure (scaling with $n_\mathrm{e}^2$) is orders of magnitude lower in the SWB than in the \ion{H}{ii} region.
It is also important to note that there is no significant density enhancement of the ISM at the edge of the SWB; this demonstrates that the SWB is not expanding supersonically and that the emission measure of the ISM is not enhanced by the SWB.

The \ion{H}{ii} region has quite different shape in HV04 compared with WV04 (see Fig.~\ref{fig:V04slice}), as traced by the contour of 50\% ionization ($x=0.5$).
The downstream part of the bubble ($z\lesssim-1.75$ pc) recombines in HV04, whereas with a stellar wind (WV04) it doesn't recombine because the density is lower and there is significant heating through turbulent mixing.
This difference could in principle be detected through the different emission measure of the two simulations, but it may be difficult because the emission measure of the wind bubble is very small.

\section{Modelling dust emission, and observational data used}
\label{sec:dust}

We model the dust emission using the Monte Carlo radiation transport and hydrodynamics code \textsc{torus} \citep{Har00, KurHarBatEA04, Har15}.

\textsc{pion} data snapshots have been saved as \textsc{fits} images and we have written a \textsc{torus} file reader to interpret these files.
For a given snapshot, the gas density, temperature, ionization state, and composition (a passive tracer distinguishing wind from ISM) are imported into \textsc{torus}.
The stellar wind from an O star contains no dust, so the composition tracer allows \textsc{torus} to ignore the SWB when performing the dust radiative transfer
(although excluding gas significantly hotter than $10^4$ K is equally effective).
The \textsc{pion} snapshots are mapped onto the \textsc{torus} grid using linear (1D) or bilinear (2D) interpolation.

For the 1D models we assume spherical symmetry and a domain somewhat larger than the \ion{H}{ii} region shell radius.
During the radiative equilibrium calculation, photon packets are propagated in 3D space, but the symmetry of the problem means that they can always be mapped back onto the 1D grid.
The 1D radiative transfer calculations use a grid of 2 pc with 512 cells, giving a zone size of $\Delta{r} = 3.906\times10^{-3}$\,pc (same as the \textsc{pion} simulations).
To make images, the image size is 2.9 pc and uses 401 pixels per side with a pixel width of $7.23\times10^{-3}$ pc.

For the 2D models we assume cylindrical symmetry about the $z$-axis, which is the star's propagation vector.
Again, photon packets are propagated through 3D space, but the symmetry of the problem means that they only ever update the 2D grid.
The radiative equilibrium calculations use a grid of size 2 pc and 512 cells per side, again matched to the \textsc{pion} simulations.
The images have a side of length 2.44 pc and 401 pixels per side, giving a pixel width of $6.08\times10^{-3}$ pc.

\subsection{Calculating dust emission with \textsc{torus}}
Dust radiative-equilibrium temperatures are computed using the \cite{Luc99} photon-packet-propagation algorithm.
The dust-to-gas ratio is assumed to be the canonical value of 0.01 \citep{DraDalBenEA07}.
The radiation source is the same as that used in the RHD simulations, a blackbody with $T_\mathrm{eff}=37\,500$ K and a luminosity normalised to give $Q_0=3\times10^{48}$ s$^{-1}$, and the whole spectrum is sampled by the radiative transfer scheme.
Only opacity from dust is considered when calculating the attenuation of radiation.
This is not strictly correct for EUV photons, but most of the radiation is emitted at lower energies, for which dust is the main opacity source.

We consider a range of different dust models, listed in Table~\ref{tab:dustmodels}, using a \cite{MatRumNor77} size distribution defined by the minimum (maximum) grain size $a_\textrm{min}$ ($a_\textrm{max}$) and a power law index $q$.
For silicates \textsc{torus} assumes spherical dust grains with optical constants taken from \cite{DraLee84}.
The different silicate models have values of $q$ ranging from 2 to 3.5, including the canonical values $q=3.3$ and 3.5 \citep[e.g.,][]{MatRumNor77}.
The \textsc{torus} model for carbon grains is the amorphous carbon grains of \citet{ZubMenColEA96}, and we only used $q=3.3$ and $q=2.0$ with these grains.
\textsc{torus} uses a pre-tabulated Mie scattering phase matrix to treat scattering.

\begin{table}
  \centering
  \caption{
    Models for dust properties used to calculate emission maps.
    The dust types are silicate \citep{DraLee84} or amorphous carbon grains \citep{ZubMenColEA96}.
    The minimum (maximum) dust size, $a_\textrm{min}$ ($a_\textrm{max}$), is measured in $\mu$m.
    $q$ is the power-law index of the grain size distribution between $a_\textrm{min}$ and $a_\textrm{max}$.
    }
  \begin{tabular}{ l c c c c}
    Model & Type & $a_\textrm{min}$ & $a_\textrm{max}$ & $q$ \\
    \hline
    Sil3.5 & silicate & 0.005  & 0.25 & 3.5  \\
    Sil3.3 & silicate & 0.005  & 0.25 & 3.3  \\
    Sil3.0 & silicate & 0.005  & 0.25 & 3.0  \\
    Sil2.0 & silicate & 0.005  & 0.25 & 2.0  \\
    AmC3.3 & carbon   & 0.005  & 0.25 & 3.3  \\
    AmC2.0 & carbon   & 0.005  & 0.25 & 2.0  \\
  \end{tabular}
  \label{tab:dustmodels}
\end{table}

Synthetic images are also calculated using Monte Carlo radiative transfer.
We use the forced first scattering technique of \citet{CasEve57} as well as the peel-off technique \citep{YusMorWhi84} to improve the signal-to-noise and reduce computation time.
Tests of this method and its application to dust around the hot Wolf-Rayet star WR104 are presented in \citet{HarMonSymEA04}.
\textsc{torus} is further tested by comparison with other radiative transfer codes in the extensive benchmark tests of \citet{PinHarMinEA09}.
We produce \textsc{fits} images of the dust emission maps, in units of MJy\,sr$^{-1}$, so that they are directly comparable to the \textit{Spitzer} and \textit{Herschel} images.

\subsection{Observational data}

Observational images of RCW\,120 were downloaded from the NASA/IPAC infrared science archive\footnote{\href{http://irsa.ipac.caltech.edu/}{http://irsa.ipac.caltech.edu/}}.
The \emph{Spitzer} 24 $\mu$m image
was observed on 2006.10.05 (Program ID 20597, P.I.:~S.~Carey), using the Multiband Imaging Photometer for \textit{Spitzer} \citep[MIPS][]{RieYouEngEA04}.
The \emph{Herschel} 70 $\mu$m image was observed on 2011.03.22 (observation IDs 1342216585 and 1342216586, P.I.:~F.~Motte).
The \emph{Herschel} 160 $\mu$m image was observed on 2009.10.09 (observation IDs 1342185553 and 1342185554, P.I.:~F.~Motte).
All images are level 2 data products, with units of MJy\,sr$^{-1}$ (or Jy\,pix$^{-1}$ which we converted to MJy\,sr$^{-1}$ using the pixel scale in the FITS header).

\section{Synthetic maps and comparison with infrared observations}
\label{sec:results}
In this section we present the dust intensity maps produced from the simulation snapshots, at wavelengths 24 $\mu$m, 70 $\mu$m, 100 $\mu$m, 160 $\mu$m, and 250 $\mu$m, corresponding to central wavelengths of \textit{Spitzer} and \textit{Herschel} broadband images.
We first discuss the spherically symmetric results (simulations HV00 and WV00), focussing on the effects of the stellar wind bubble and of different dust grain sizes and compositions.
We then show the 2D images for the moving star simulations, HV04 and WV04, and compare the emission maps with observations.
Finally we quantitatively compare the emission from simulations WV04 and HV04 with observations as a function of distance from the ionizing star.

\subsection{Static star models, WV00 and HV00}

\begin{figure}
\centering
\includegraphics[width=0.85\hsize]{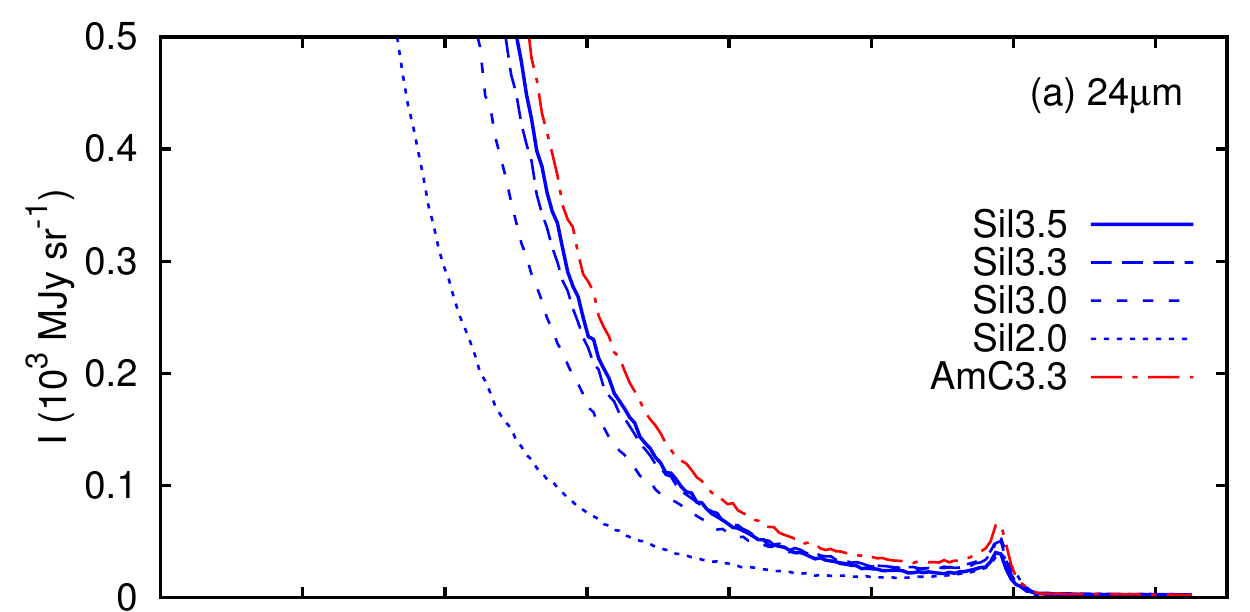}
\includegraphics[width=0.85\hsize]{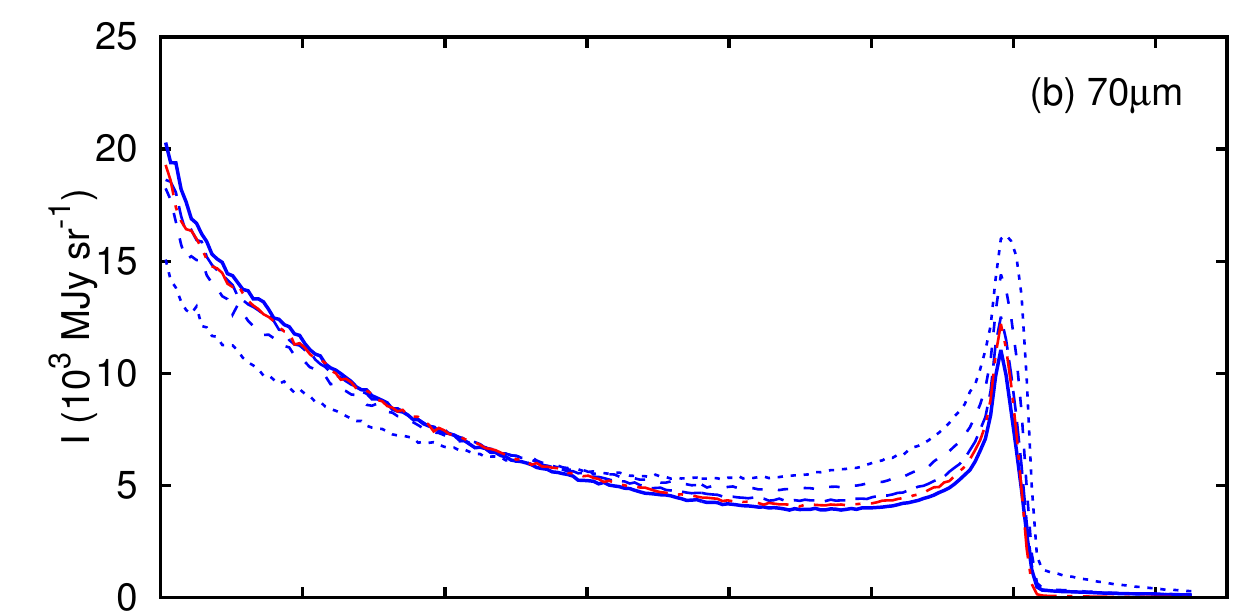}
\includegraphics[width=0.85\hsize]{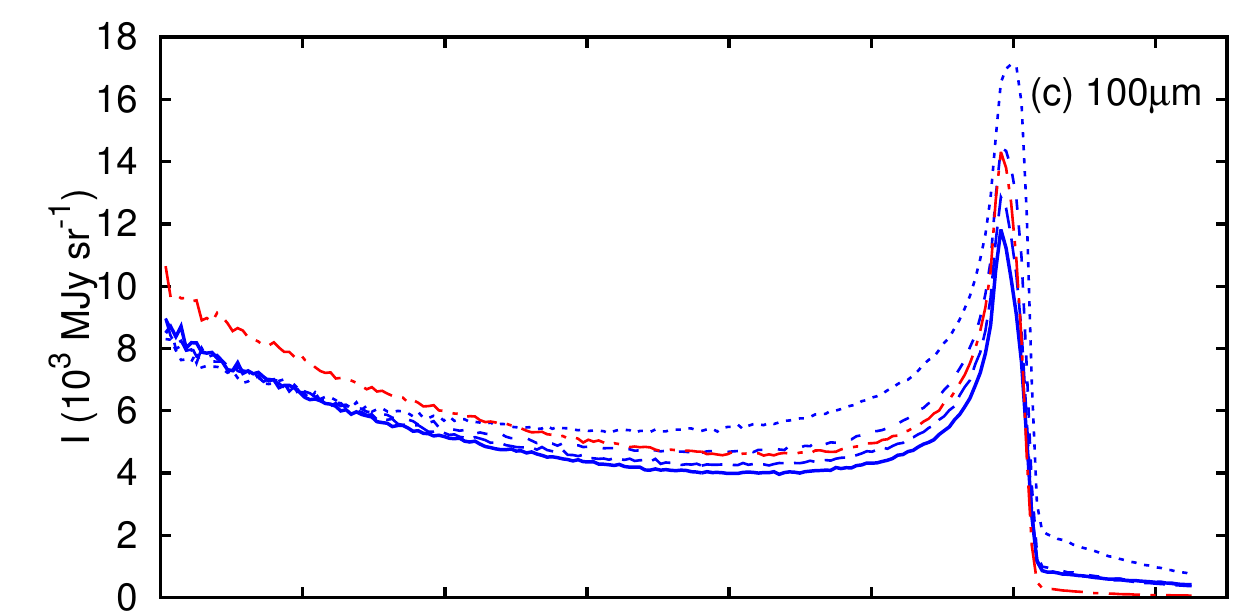}
\includegraphics[width=0.85\hsize]{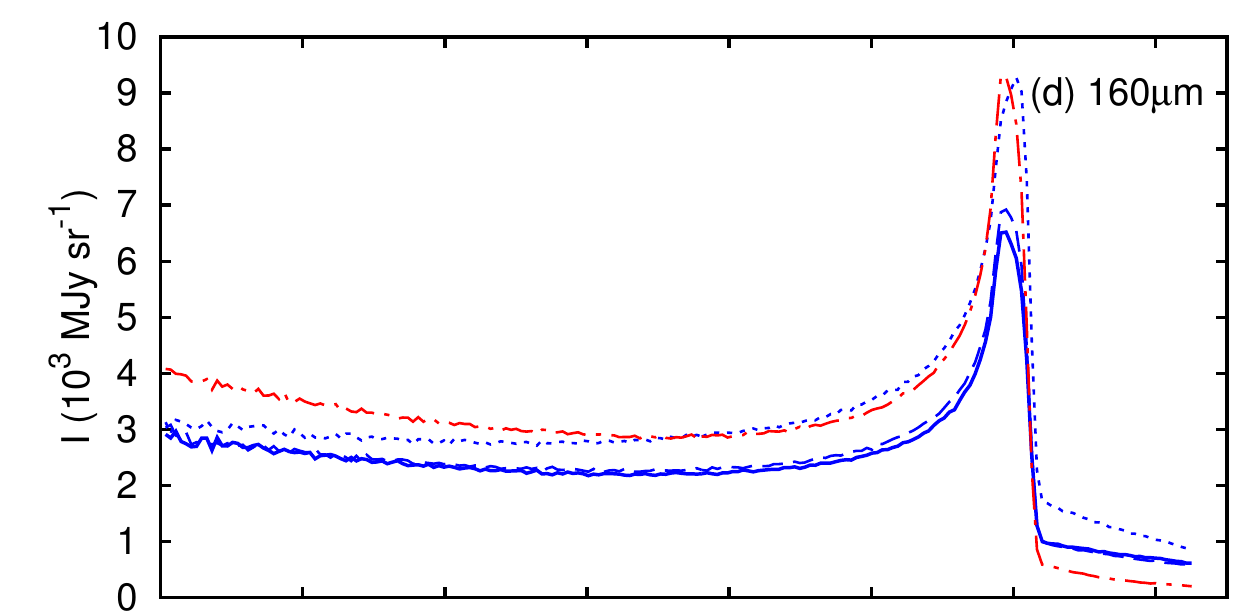}
\includegraphics[width=0.85\hsize]{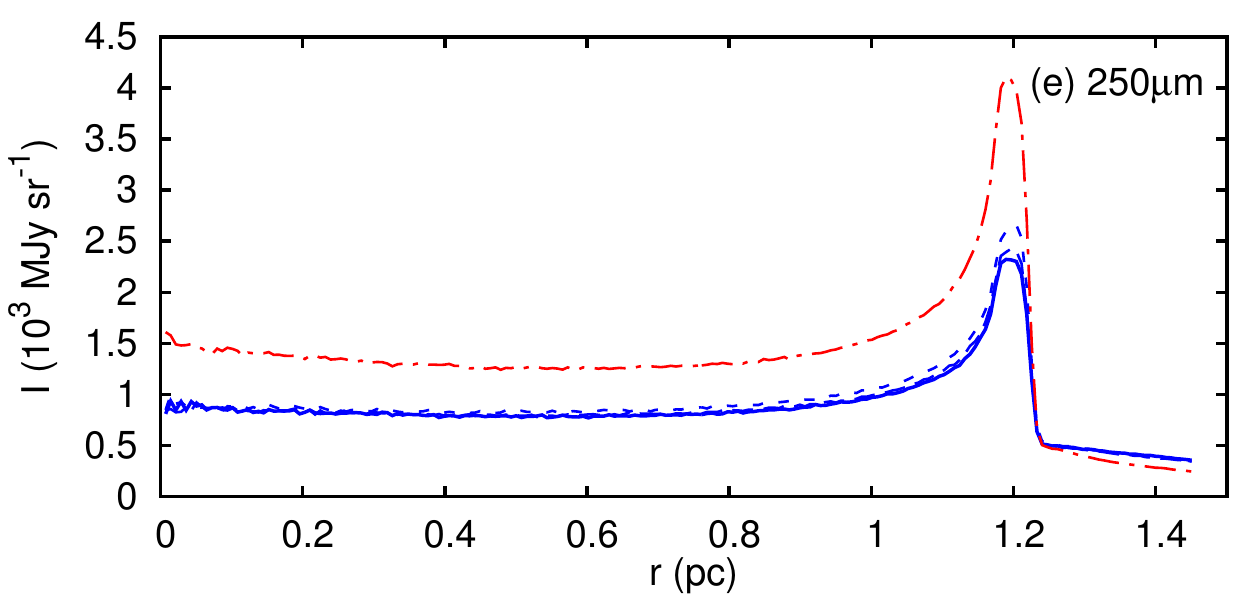}
\caption{
   Dust emission (spherically averaged) for the 1D simulation H00 at time 0.2 Myr, using different grain size distributions and dust models.
  Panels show (a) 24 $\mu$m emission, (b) 70 $\mu$m, (c) 100 $\mu$m, (d) 160 $\mu$m, and (e) 250  $\mu$m.
  "Sil" is for silicate grains, "AmC" is for amorphous carbon grains, and the number is the power law index, $q$, of the grain size distribution, as listed in Table~\ref{tab:dustmodels}.
  We plot the intensity (or surface brightness, measured in MJy\,sr$^{-1}$) as a function of distance, $r$, from the ionizing star (in parsecs).
  There is no wind bubble, so the only peaks are at $r=0$ and the \ion{H}{ii} region shell at $\approx1.2$ pc.
  }
\label{fig:dustHV00}
\end{figure}

\begin{figure}
\centering
\includegraphics[width=0.85\hsize]{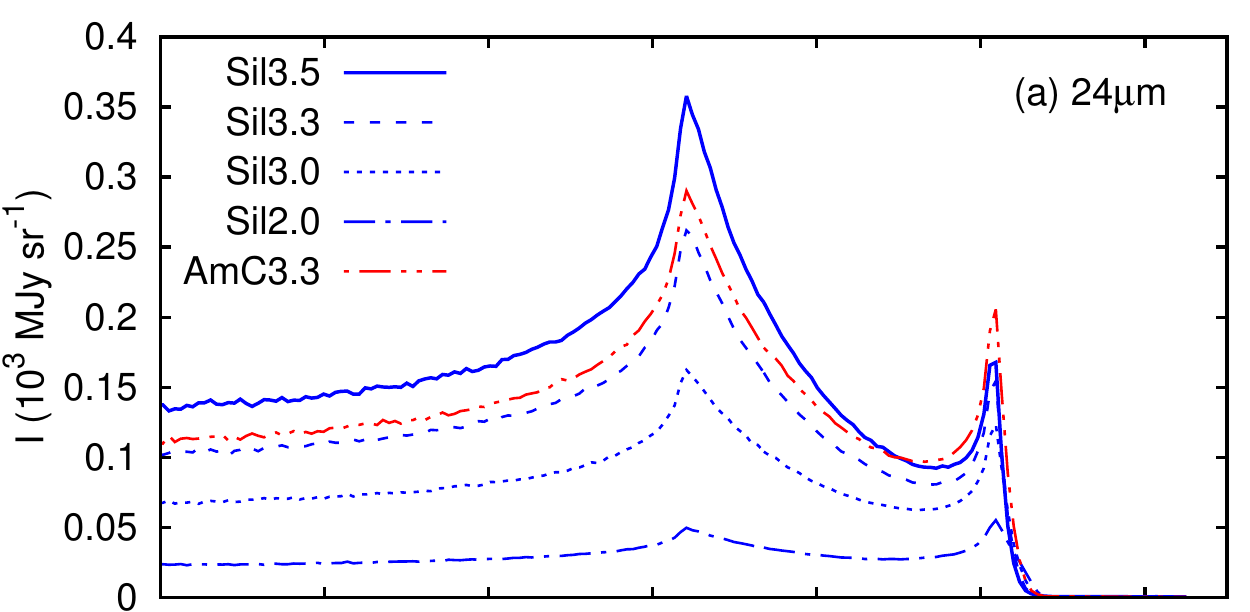}
\includegraphics[width=0.85\hsize]{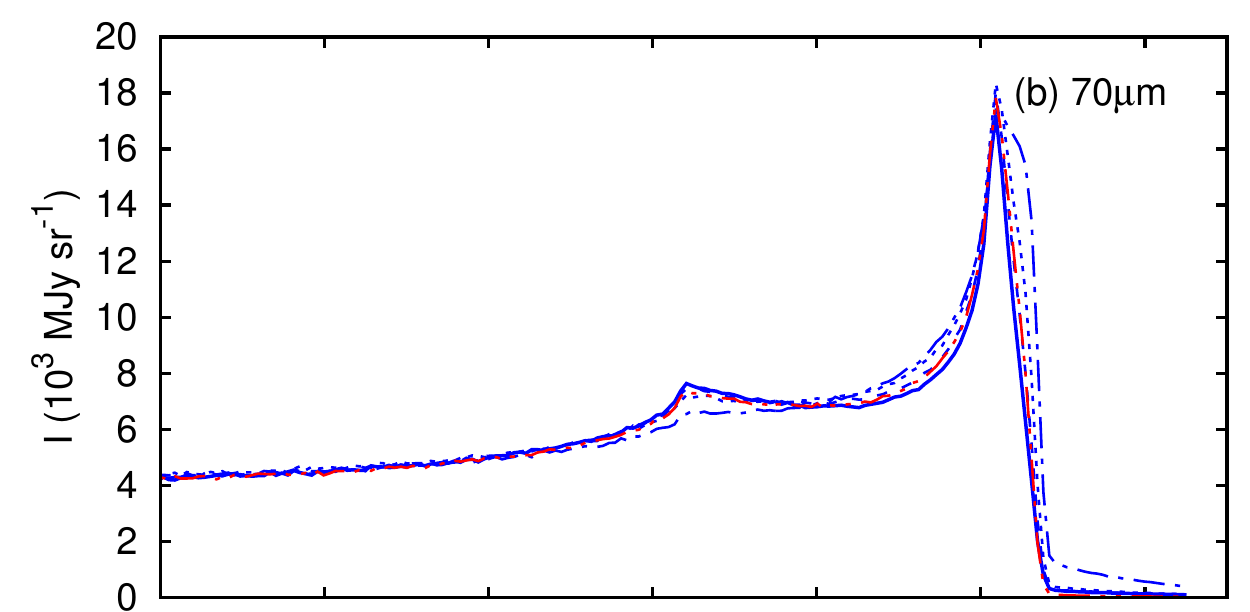}
\includegraphics[width=0.85\hsize]{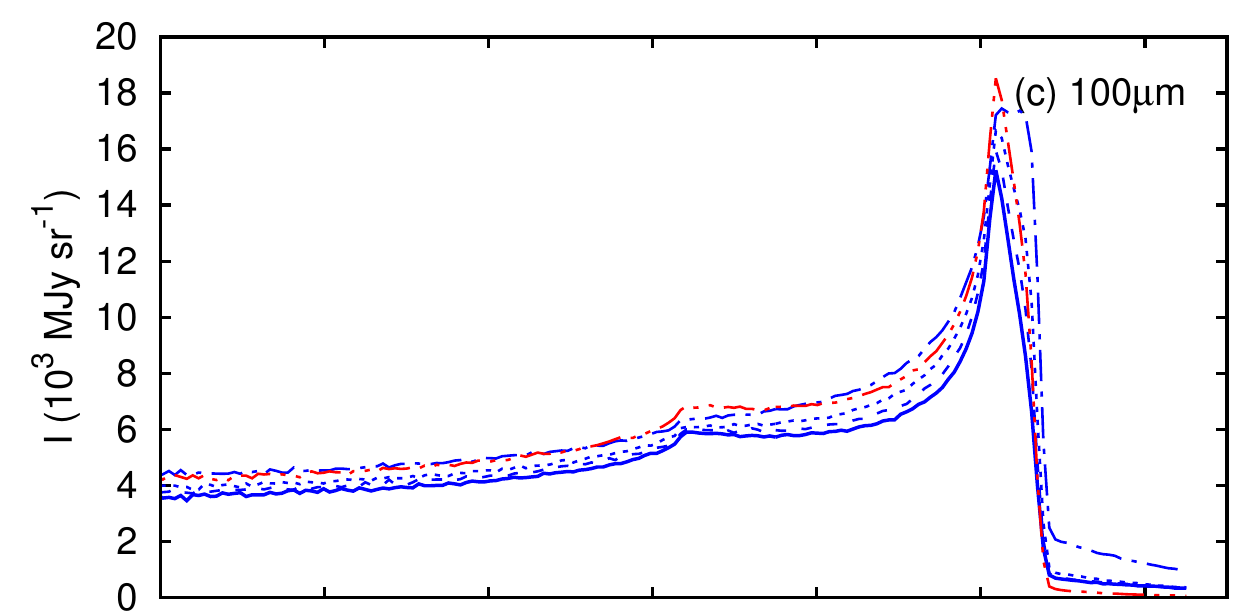}
\includegraphics[width=0.85\hsize]{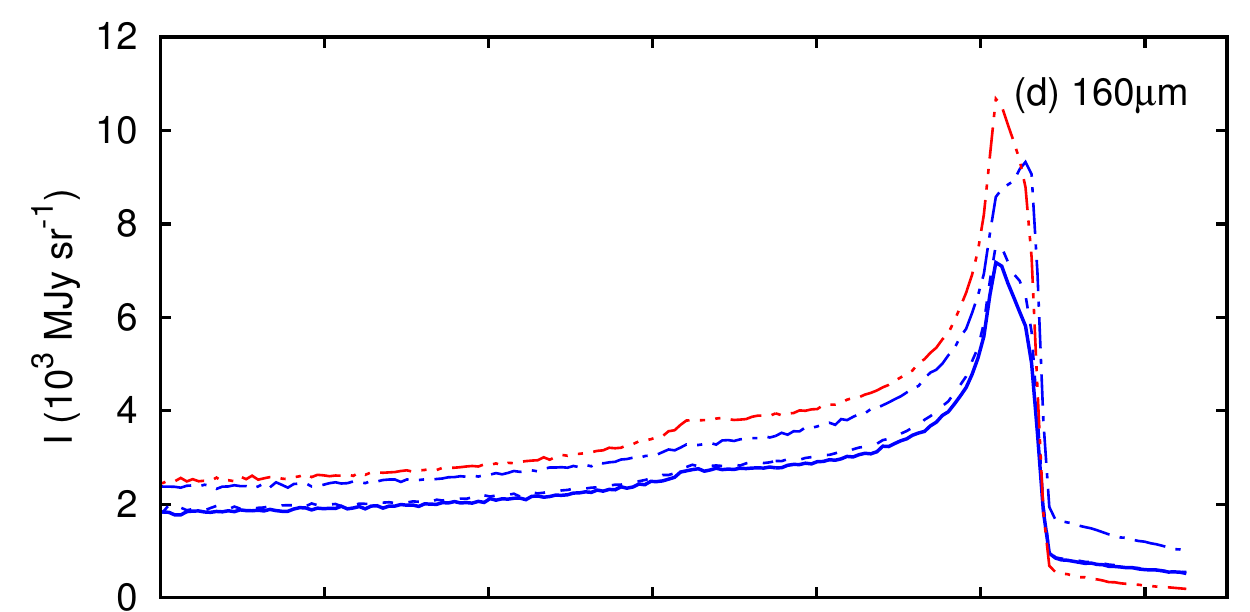}
\includegraphics[width=0.85\hsize]{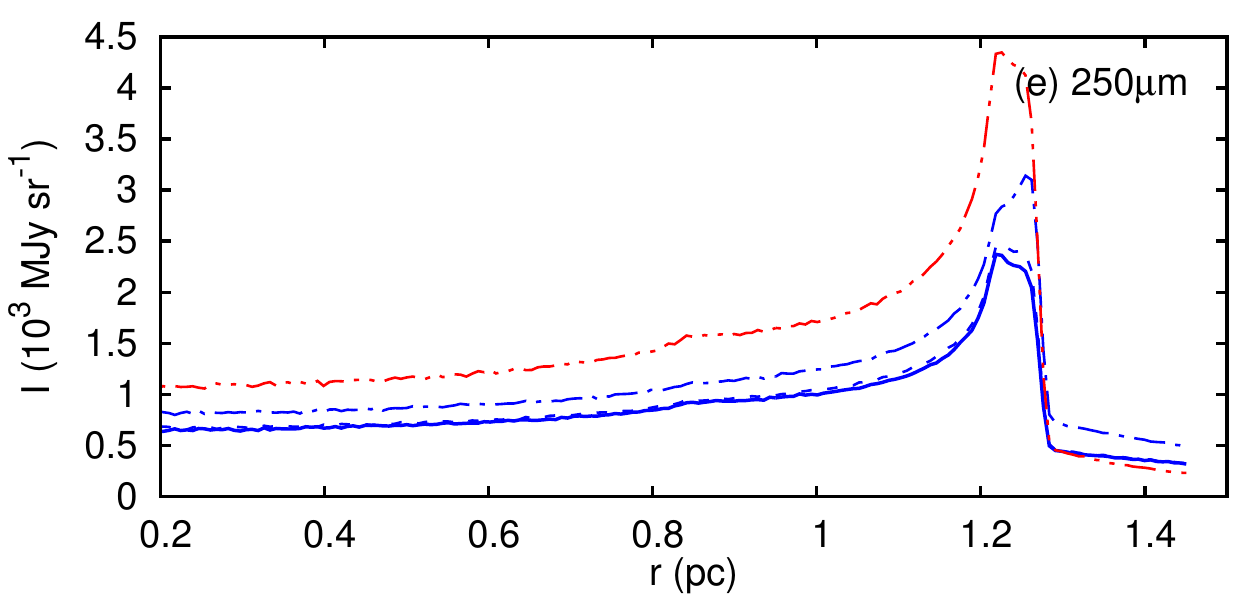}
\caption{
  Dust emission (spherically averaged) for the 1D simulation W00 at time 0.2 Myr, using different grain size distributions and dust models.
  Panels show (a) 24 $\mu$m emission, (b) 70 $\mu$m, (c) 100 $\mu$m, (d) 160 $\mu$m, and (e) 250  $\mu$m.
  We plot the intensity (or surface brightness, measured in MJy\,sr$^{-1}$) as a function of distance, $r$, from the ionizing star (in parsecs).
  The inner peak at $\approx0.85$ pc traces the edge of the wind bubble, and the outer peak at $\approx1.2$ pc traces the \ion{H}{ii} region shell.
  }
\label{fig:dustWV00}
\end{figure}

For the spherically symmetric calculations we used \textsc{torus} to experiment with a number of different grain size distributions and properties, listed in Table~\ref{tab:dustmodels}.
The brightness of the projected dust emission as a function of distance from the central star is plotted for simulation HV00 in Fig.~(\ref{fig:dustHV00}), and for WV00 in Fig.~(\ref{fig:dustWV00}).

\subsubsection{Intense emission at 24 $\mu$m}
At 24 $\mu$m there is a clear difference between the predictions of the two simulations, in that HV00 has much brighter emission at small radii than WV00.
This is because HV00 has an almost uniform density \ion{H}{ii} region (containing dust) that extends all the way to the star, whereas WV00 has a stellar wind bubble (a cavity of very low-density and dust-free gas) extending from the star to $r\approx0.8$ pc.
In model HV00 the dust close to the star is very warm and so emits incredibly brightly.
The brightness of emission for HV00 increases towards $r=0$ at all wavelengths, but this increase becomes much stronger at shorter wavelengths.
This makes sense because the $24\ \mu$m emission is from the Wien tail of the dust blackbody spectrum for almost all dust temperatures in \ion{H}{ii} regions (e.g.,\ at dust temperature $T_\mathrm{d}=50$ K, the peak of the blackbody spectrum is at 58 $\mu$m), so a small increase in temperature can lead to a large increase in emissivity.

\begin{figure}
\centering
\includegraphics[width=0.85\hsize]{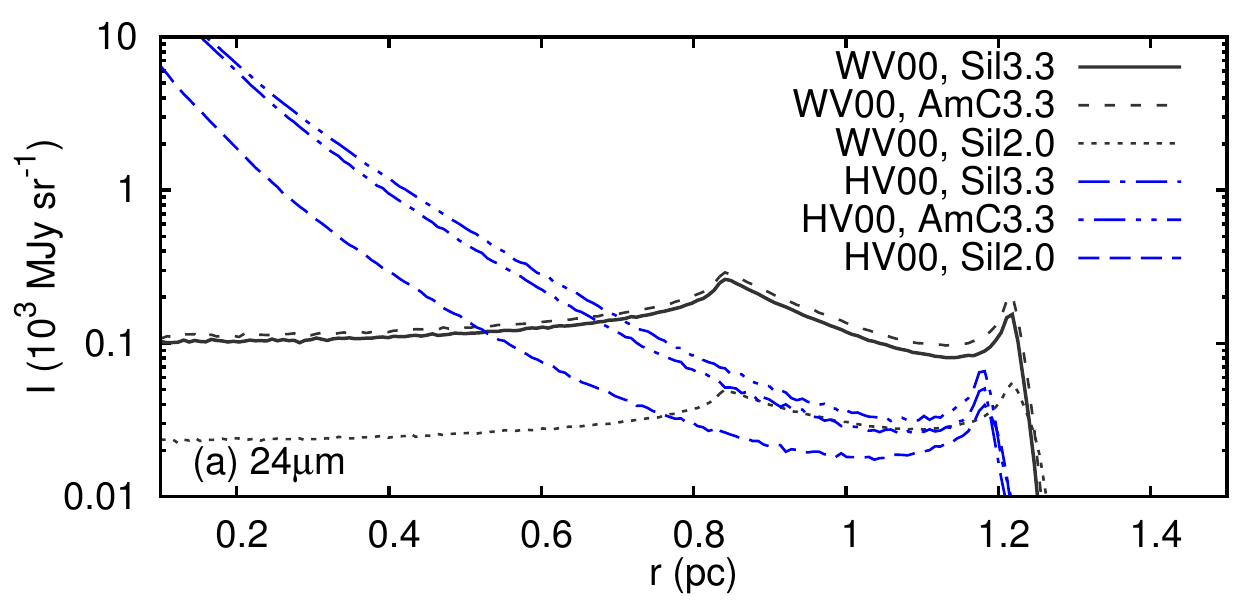}
\caption{
  Dust emission (spherically averaged) at 24 $\mu$m for the 1D simulations WV00 and HV00 at time 0.2 Myr, using different grain size distributions and dust models.
  We plot the intensity (or surface brightness, measured in MJy\,sr$^{-1}$) as a function of distance, $r$, from the ionizing star (in parsecs), with a logarithmic $y$-axis to show the exponential decrease in brightness with radius.
  }
\label{fig:compV00}
\end{figure}

Fig.~\ref{fig:compV00} shows a comparison between dust emission from HV00 and WV00 at $\lambda=24$ $\mu$m, but this time using a logarithmic $y$-axis to show the full range in brightness.
We see that the brightness decreases approximately exponentially for HV00 from $r\approx(0.1-0.8)$ pc, which can be attributed to the steady decrease in $T_\mathrm{d}$ with increasing radius.
The same decrease is seen for WV00 from the outer edge of the wind bubble (0.85 pc) to near the \ion{H}{ii} region border (1.2 pc), until the projected contribution from the dense shell becomes comparable to the interior emission.

\subsubsection{Outer shell emission at 250 $\mu$m}
At 250 $\mu$m, by contrast, the emission comes from the Rayleigh-Jeans part of the spectrum, where emissivity is linearly proportional to dust temperature (and density), and so we see a much smaller increase towards $r=0$.
The shell is 200 times denser than the \ion{H}{ii} region whereas its temperature is only a factor of 2-3 smaller, and so shell emission is up to 100 times brighter than interior emission.
This means that the projected emission in Fig.~\ref{fig:dustHV00}(e) is almost entirely from the shell at all radii, with just the slight increase as $r\rightarrow0$ being due to interior emission.
This is also true to a lesser extent for 160 $\mu$m emission in Fig.~\ref{fig:dustHV00}(d).

\subsubsection{Intermediate wavelengths}
At 70 and 100 $\mu$m, the emission from HV00 peaks at the origin and within the dense neutral shell at the \ion{H}{ii} region border, with comparable brightness at both peaks.
The inner peak is a dust temperature effect, with hotter dust near $r=0$ emitting more than the cooler dust at larger radii.
The outer peak is a density effect: even though dust in the shell is cooler than in the \ion{H}{ii} region, there is much more dust, so it emits brightly.
For WV00, Fig.~\ref{fig:dustWV00} shows that the SWB is still visible at 70 and 100 $\mu$m, although not as bright as at 24 $\mu$m, and that at longer wavelengths it becomes very difficult to detect.

\subsubsection{Effects of dust composition}
At short wavelengths the difference between AmC3.3 and Sil3.3 is minimal, but the difference between e.g.,\ Sil3.3 and Sil2.0 is large.
At 24 $\mu$m the models with more small grains (larger $q$) are brighter, which can be understood quite simply.
Because of their absorption and emission properties, big grains exposed to UV radiation are cooler than small grains, and so the grains within the \ion{H}{ii} region are 5-10\% cooler for Sil2.0 than for Sil3.3.
This means that at 24 $\mu$m (in the Wien tail of the spectrum), the Sil2.0 model produces about 7 times less emission than Sil3.3.

At longer wavelengths, by contrast, we see that the shell emits more brightly with Sil2.0 than for Sil3.3.
The big grains (Sil2.0) have a weaker peak at $r=0$ for HV00 (at 24 and 70 $\mu$m) because the grains are marginally cooler at small radii.
They have a stronger peak at the dense shell, however, and again it is because the big grains absorb less than the small grains.
This means that the FUV radiation penetrates deeper into the dense shell for Sil2.0 than for Sil3.3, so that dust in the outer parts of the shell is warmer for Sil2.0 than for Sil3.3.
The larger FUV irradiation more than compensates for the smaller cross-section of the bigger grains.

The model with AmC3.3 grains is brighter than Sil3.3 at long wavelengths (by nearly a factor of 2 at 250 $\mu$m); this is because of the different optical properties of the grains.
They absorb more FUV flux in the shell than the Sil3.3 model (and also even more than Sil2.0), and so more energy is re-emitted.

\subsubsection{Visibility of the wind bubble}
The most obvious comparison between HV00 and WV00 is that the wind bubble is so clearly detected at 24 $\mu$m.
This is because the emissivity of the \ion{H}{ii} region gas at 24 $\mu$m is larger at $r\approx0.8$ pc than it is in the dense shell (the temperature effect discussed above), and because the difference depends exponentially on radius.
At longer wavelengths, where the emissivity depends only linearly on dust temperature, the emission is dominated by the neutral shell and the wind bubble can only be detected by a small decrement in emission at small radii.
This could be detected at 70 and 100 $\mu$m with good data, but at longer wavelengths the wind bubble becomes more difficult to see.
H$\alpha$ line emission and radio bremsstrahlung are also proportional to $n_\mathrm{e}^2$ with weak dependence on $T$, hence wind bubbles can only be easily detected with these tracers if the bubble is expanding supersonically \citep[e.g.,~NGC\,7635;][]{ChrGouMeaEA95} or through the decrement in emission from the cavity \citep[e.g.,][]{WatPovChuEA08}.

\begin{figure}
\centering
\includegraphics[width=0.9\hsize]{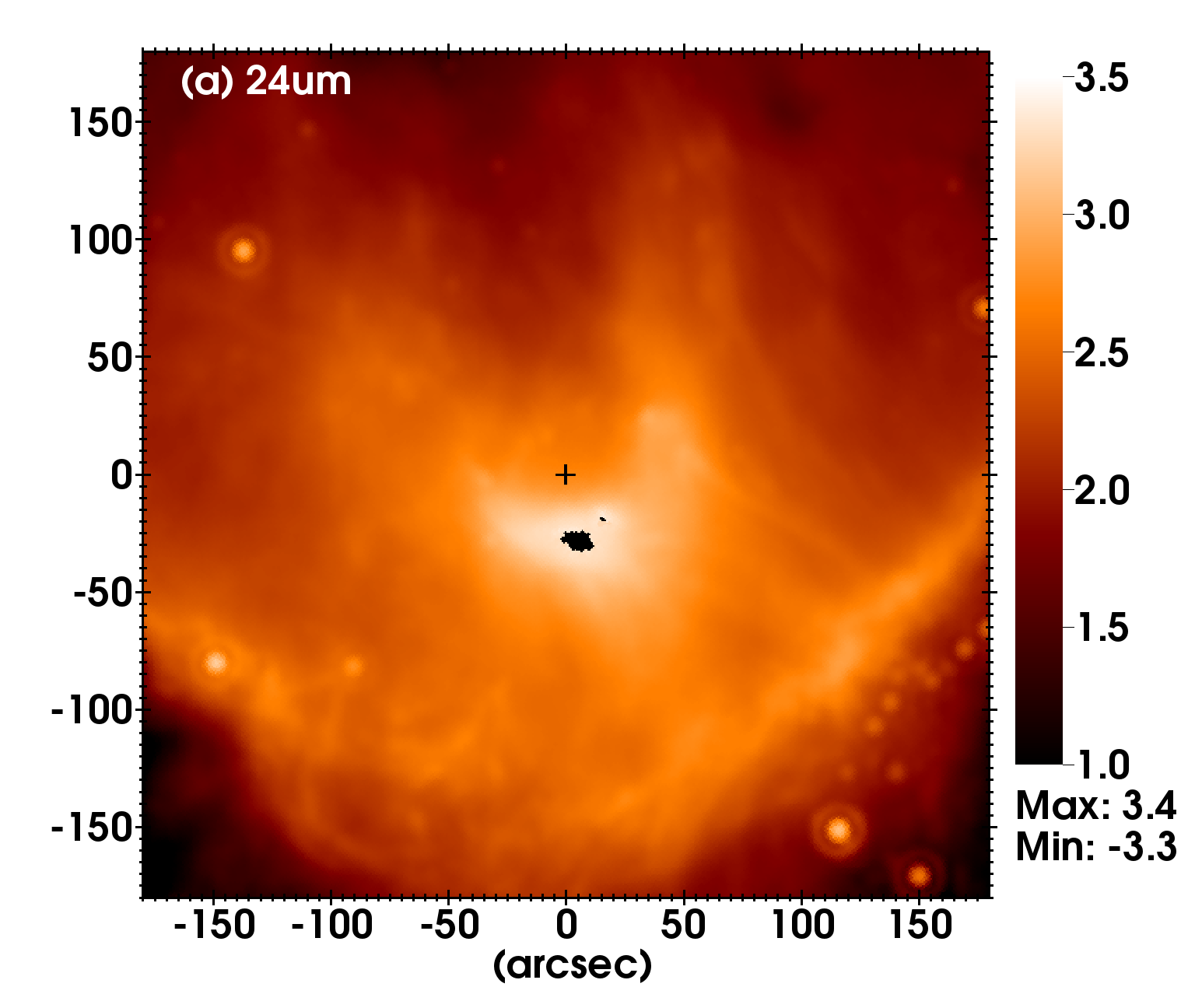}
\includegraphics[width=0.9\hsize]{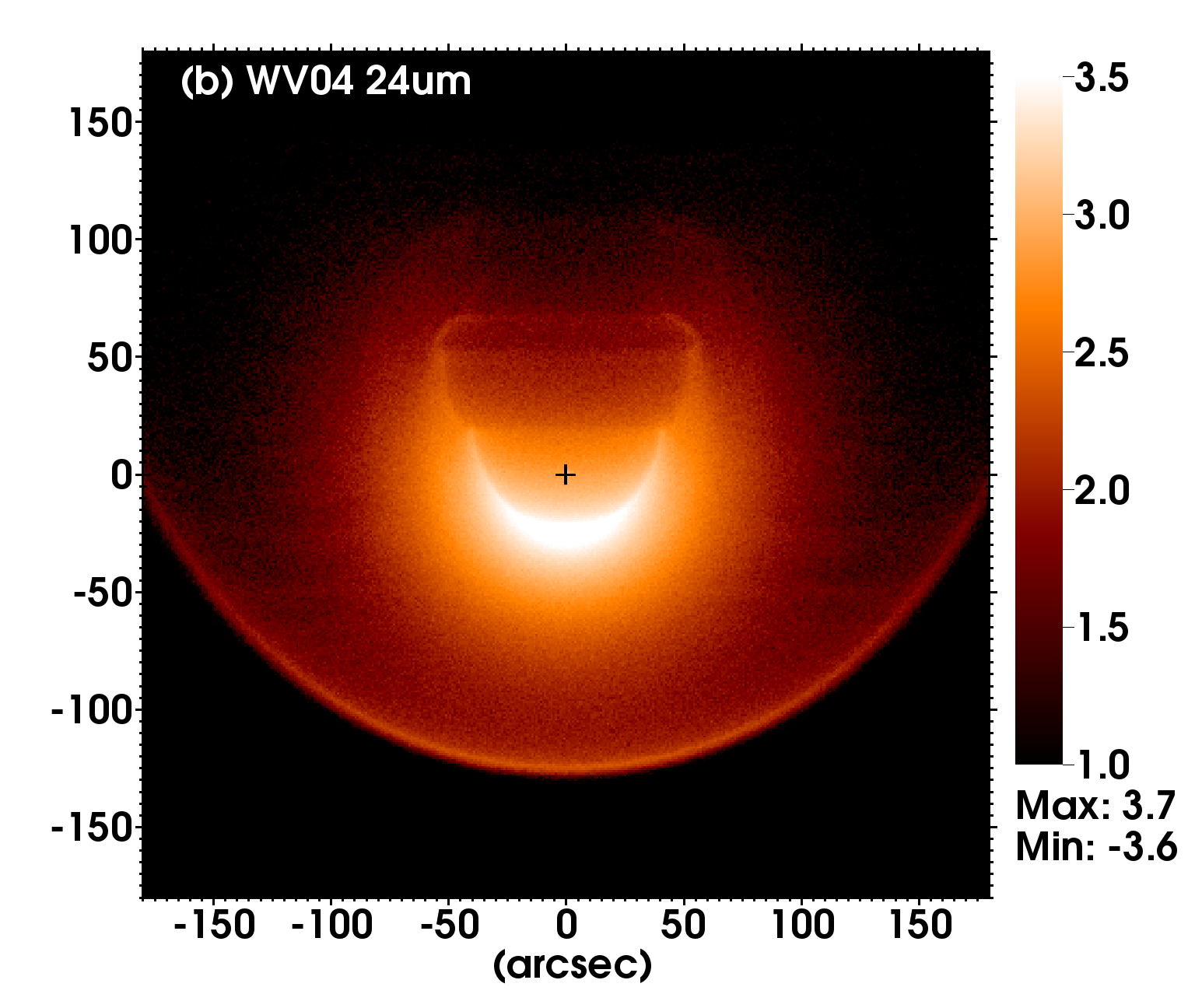}
\includegraphics[width=0.9\hsize]{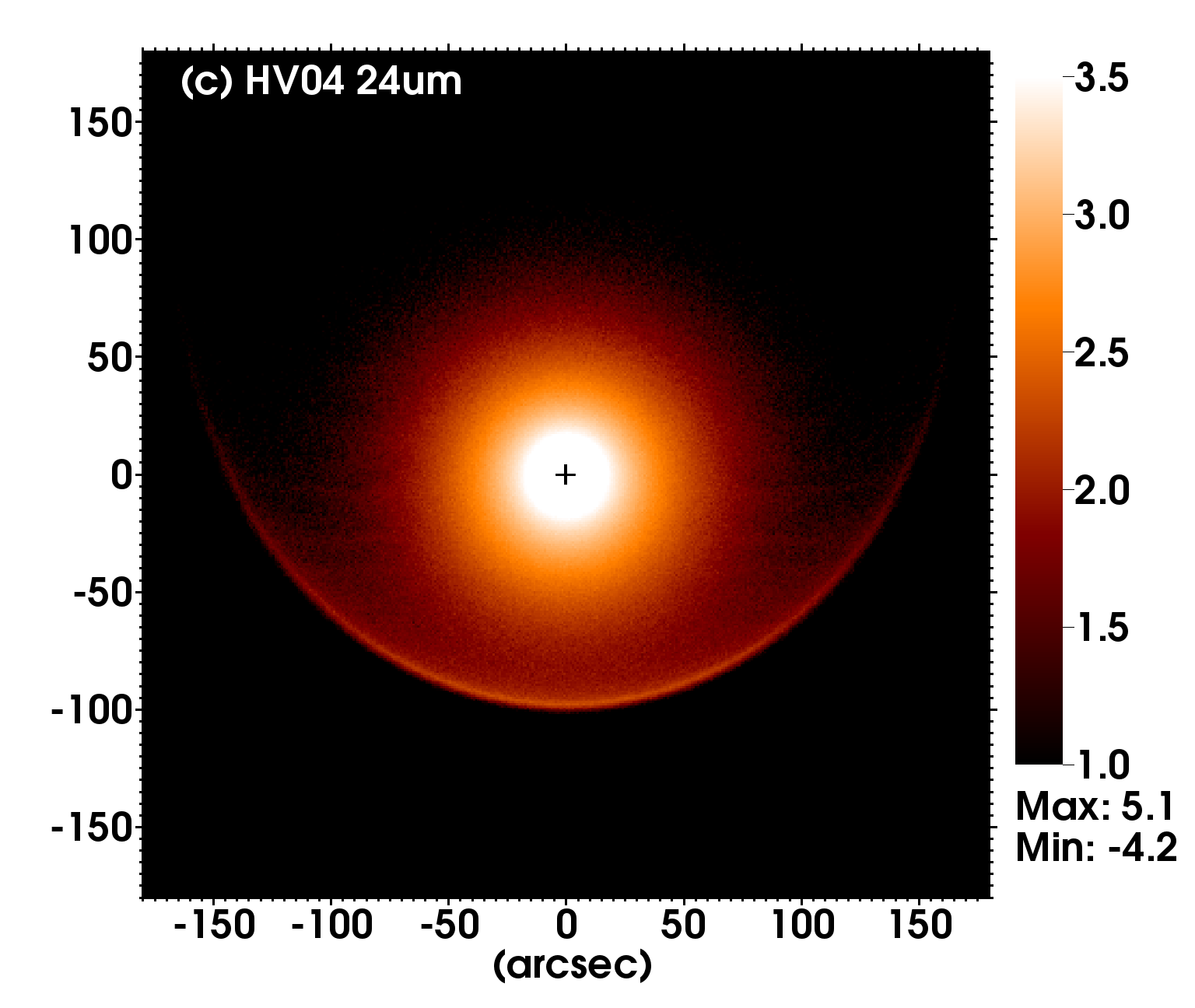}
\caption{
  \textbf{(a)} The Galactic \ion{H}{ii} region RCW\,120 in the mid-IR from \emph{Spitzer} at 24 $\mu$m.
  A dark spot below the position of the star is from saturated pixels.
  \textbf{(b)} Synthetic observation at 24 $\mu$m of dust emission from the  simulation WV04 (sil2.0).
  \textbf{(c)} Synthetic observation at 24 $\mu$m of dust emission from the  simulation HV04 (sil2.0).
  The black cross shows the position of the ionizing star.
  The coordinates are in arcseconds relative to the star's position.
  The units are in MJy\,sr$^{-1}$, shown in logarithmic units (i.e.~$\log_{10} I$).
  The outer arc/ring is the \ion{H}{ii} region boundary and the inner arc may show the edge of the stellar wind bubble.
  Images are generated from the simulations after 0.4\,Myr of evolution in this and the following figures (see Fig.~\ref{fig:V04slice}).
  }
\label{fig:rcw120_24um}
\end{figure}

\begin{figure}[h]
\centering
\includegraphics[width=0.9\hsize]{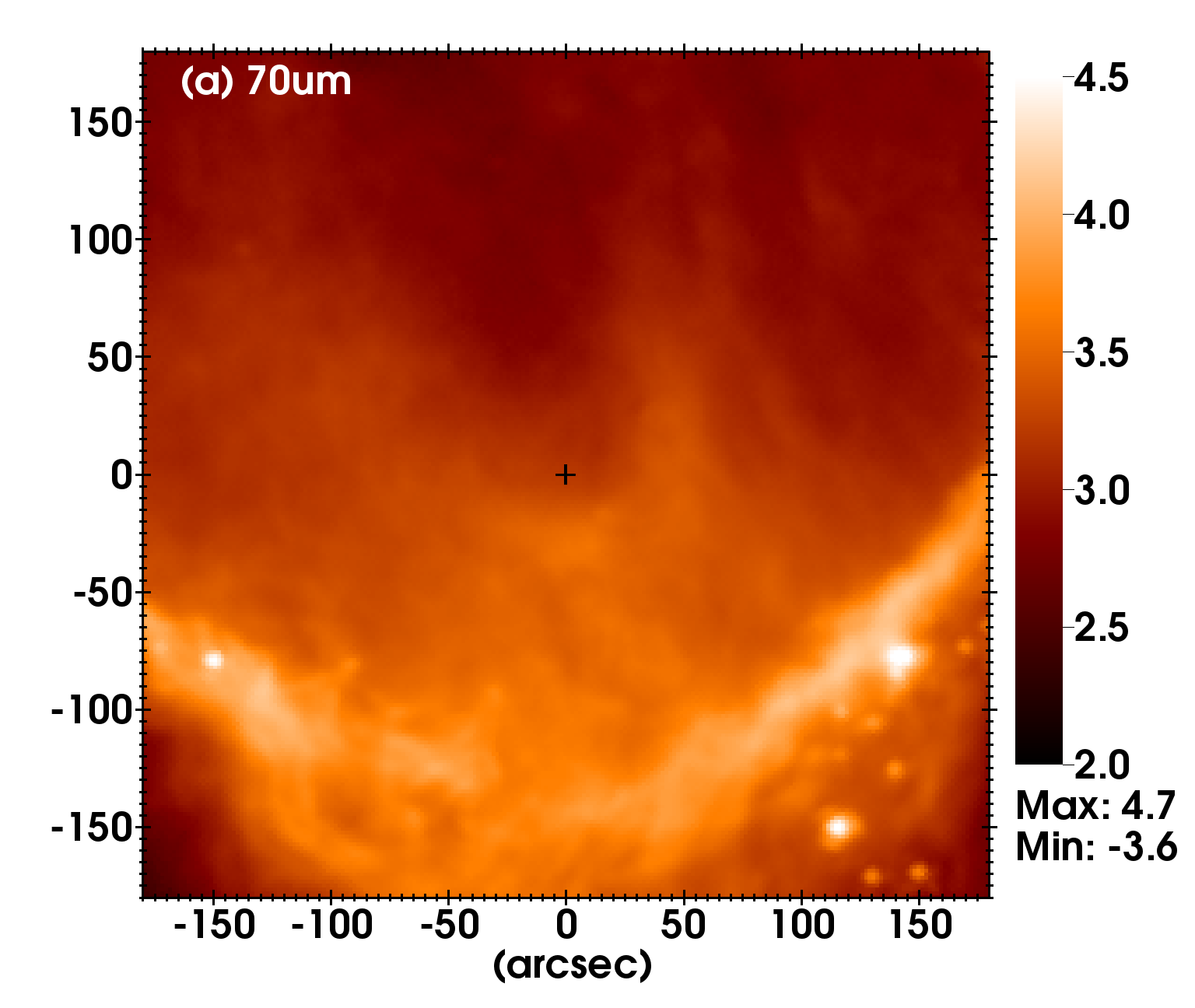}
\includegraphics[width=0.9\hsize]{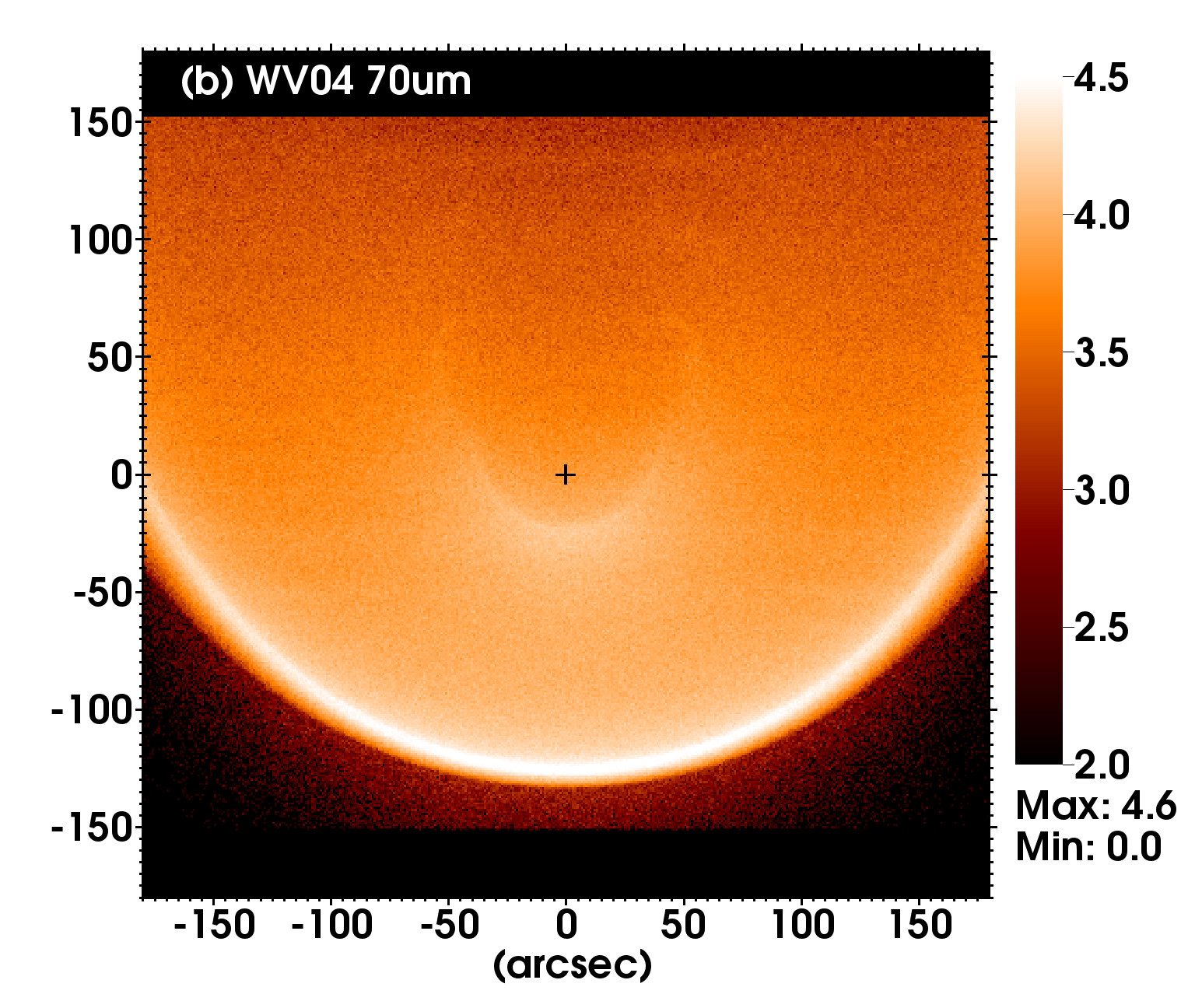}
\includegraphics[width=0.9\hsize]{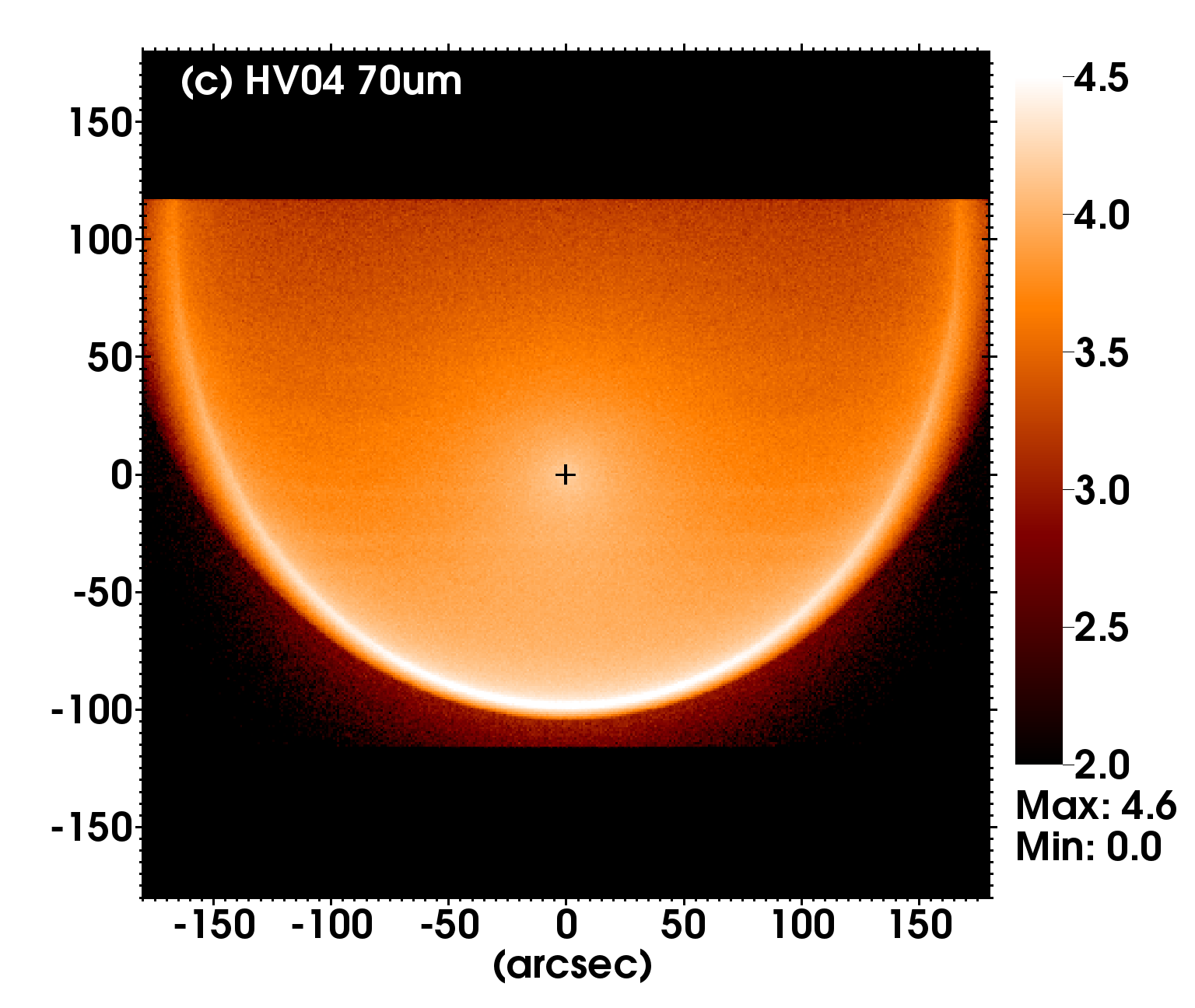}
\caption{
  As Fig.~\ref{fig:rcw120_24um}, but at 70 $\mu$m, using observational data from \emph{Herschel}.
  }
\label{fig:rcw120_70um}
\end{figure}

\begin{figure}
\centering
\includegraphics[width=0.9\hsize]{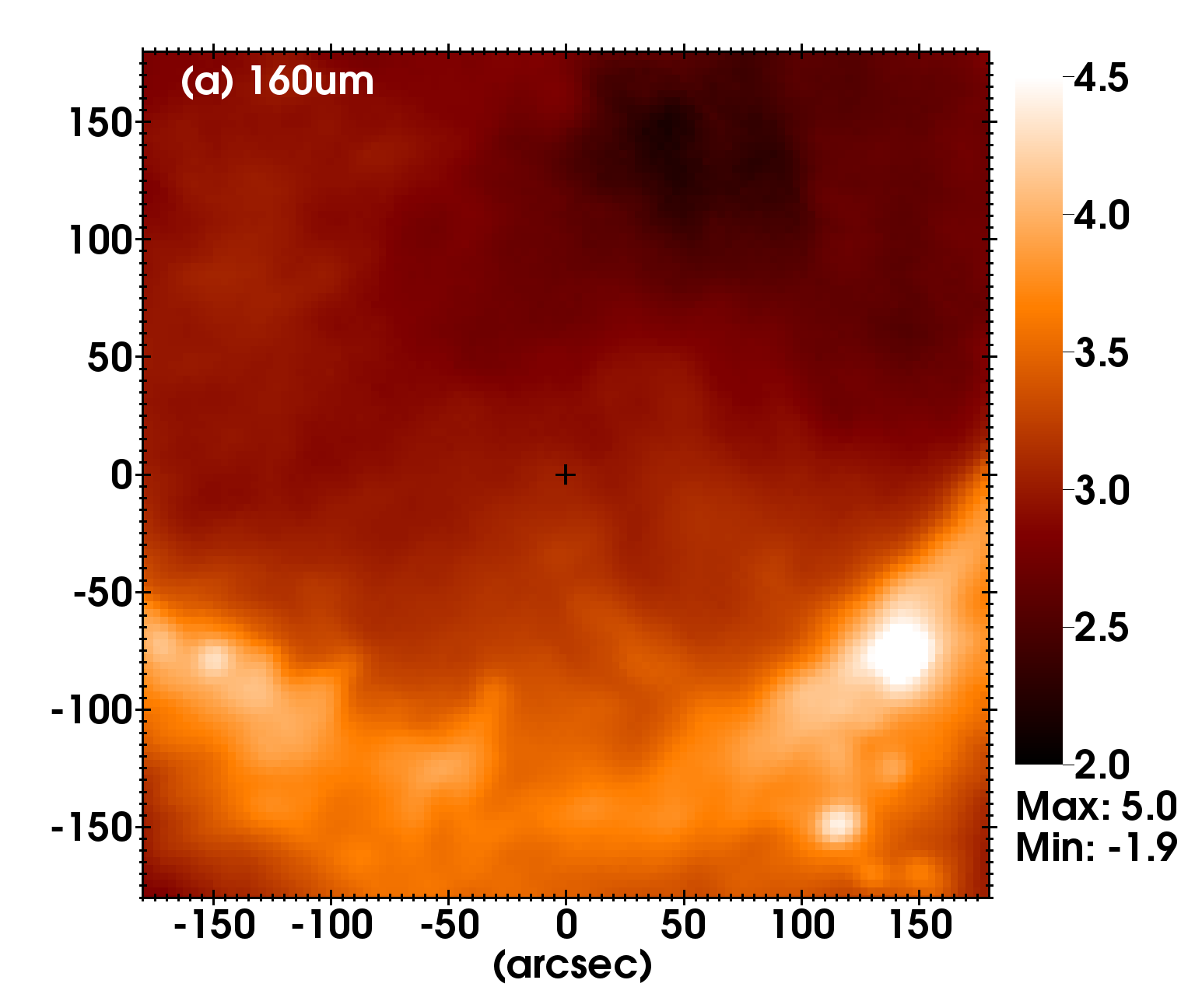}
\includegraphics[width=0.9\hsize]{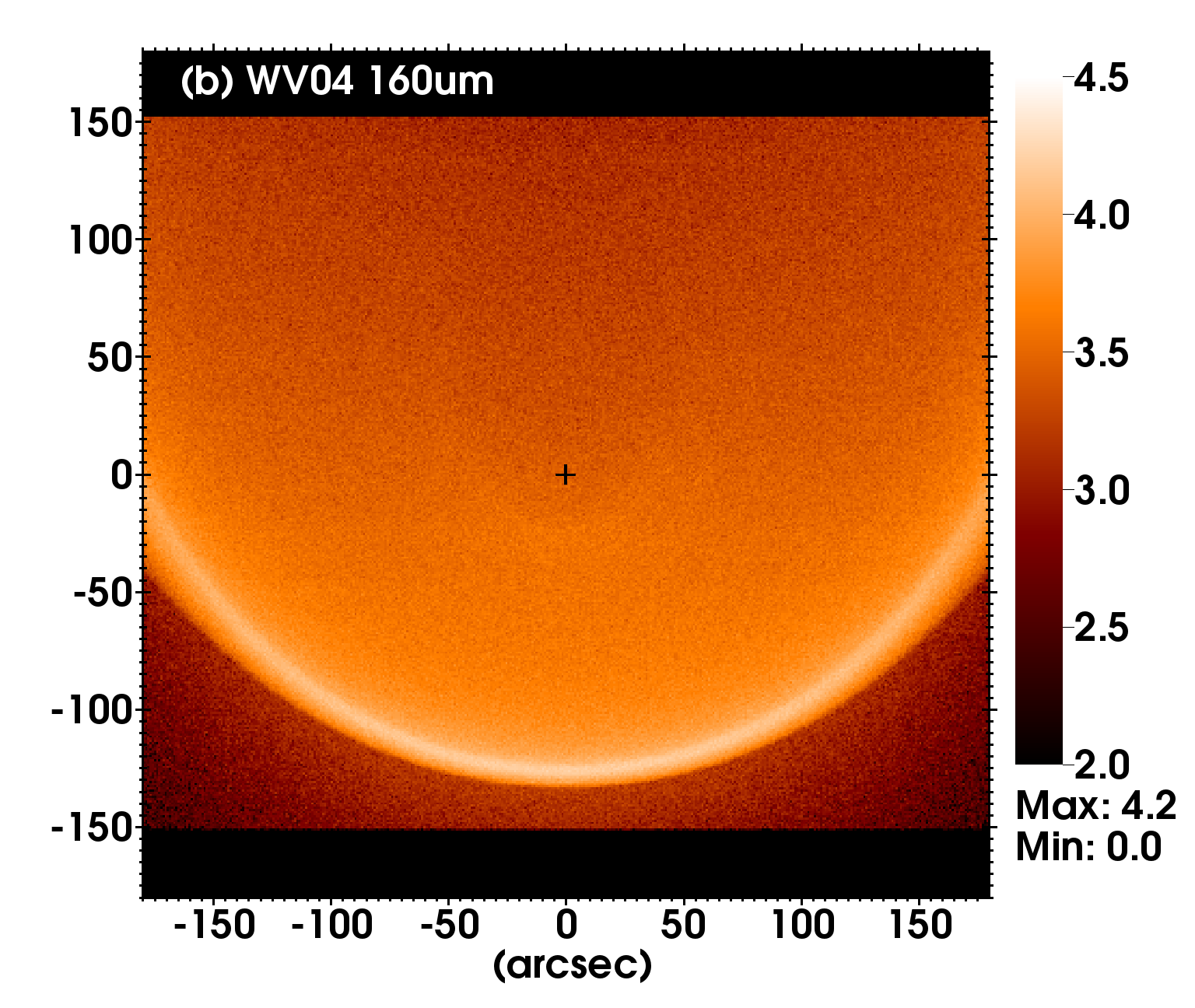}
\includegraphics[width=0.9\hsize]{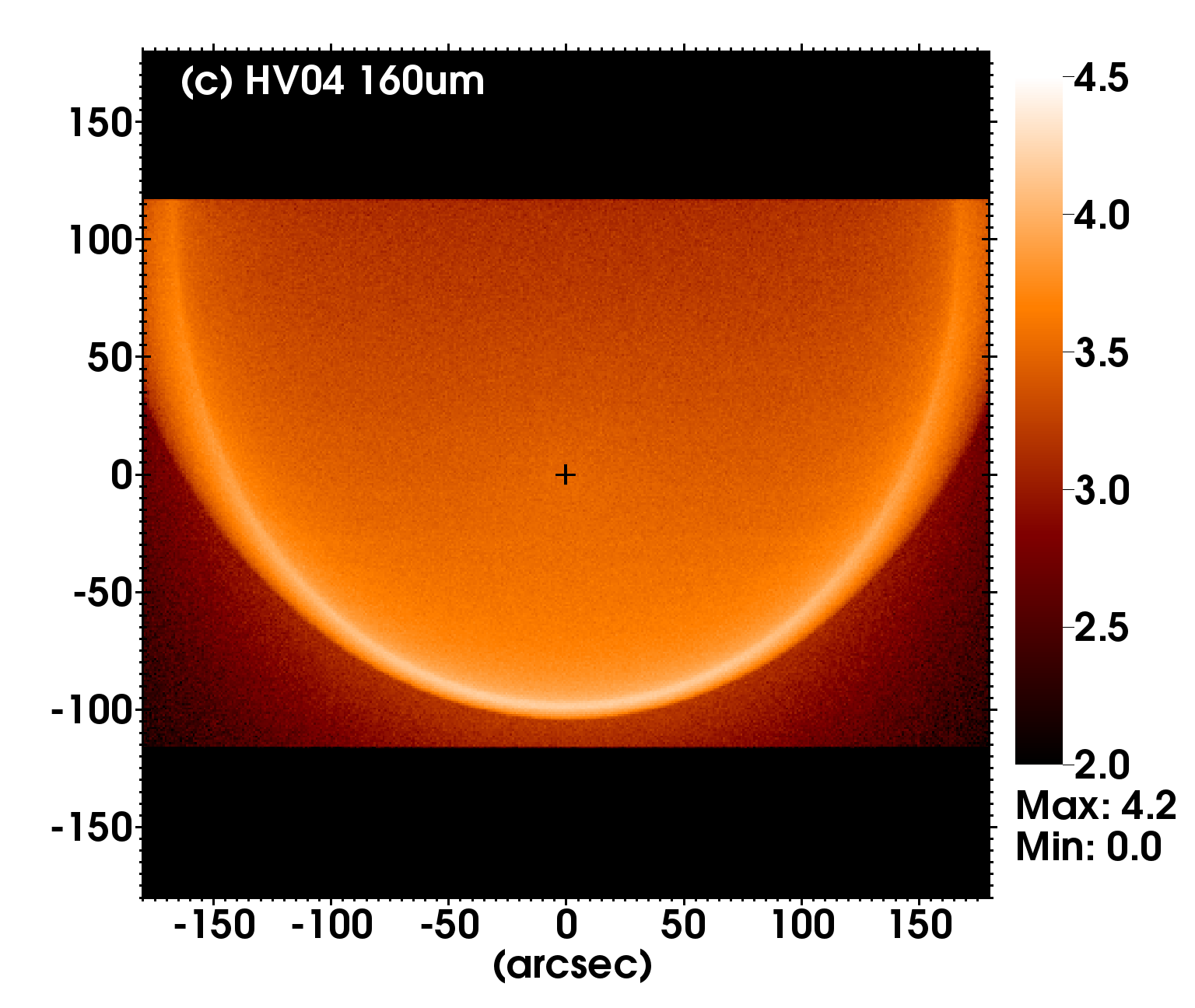}
\caption{
  As Fig.~\ref{fig:rcw120_24um}, but at 160 $\mu$m, using observational data from \emph{Herschel}.
  }
\label{fig:rcw120_160um}
\end{figure}

\subsection{Observations and synthetic images}
\label{sec:synthetic_images}

Fig.~\ref{fig:rcw120_24um} shows a comparison of synthetic 24 $\mu$m emission maps from simulations WV04 and HV04 (using dust model sil2.0) with the \emph{Spitzer} observational data.
The synthetic images are projected perpendicular to the axis of symmetry (and direction of motion of the star).
Figs.~\ref{fig:rcw120_70um} and \ref{fig:rcw120_160um} show the same comparison but at 70 $\mu$m and 160 $\mu$m, respectively, and compared with \emph{Herschel} observational data.
There are definite morphological similarities between the simulation WV04 and the observations, most obviously the arc of emission surrounding the stellar wind bubble.
This arc is very bright at 24 $\mu$m, clearly visible at 70 $\mu$m, and progressively more difficult to see at longer wavelengths.
The \ion{H}{ii} region boundary is also visible as the larger arc, where the FUV radiation from the ionizing star is heating the \ion{H}{ii} region shell.
HV04 does not fit the data well (as expected) because it has no means of creating a dust-free zone near the star.
From this we can conclude that there is a dust-free region around CD\,$-$38$\degr$11636, and that this region is about the size of the simulated stellar wind bubble.
This shows that the 24 $\mu$m emission arc is (at least qualitatively) compatible with the interpretation that it shows the edge of an asymmetric stellar wind bubble.

There are also, however, notable differences between WV04 and the observations at 24 $\mu$m.
The inner arc is somewhat brighter in the synthetic images (6300 MJy\,sr$^{-1}$) than in the observational image (2200 MJy\,sr$^{-1}$), although the brightest pixels in the observations are saturated so we do not know what the true peak brightness is.
Furthermore, the gap between the inner and outer arcs has weaker emission in the synthetic images, possibly because the outer shell is broken and broad in the observations, in contrast to the relatively thin and smooth simulated shell.
The outer arc is thinner, fainter, and smoother in the synthetic images.
This is partly by design: we set a large minimum temperature in the neutral gas to prevent fragmentation (see Paper I), but it still seems that much more shell emission is observed than is predicted.
This could be because RCW\,120 is not axisymmetric, or perhaps that the axis of symmetry is not perpendicular to the line of sight (effects of different orientations are discussed below).
It could also be because much of the mid-IR emission at 24 $\mu$m is thought to be produced by very small grains that are not in thermal equilibrium but rather are stochastically heated by photon absorptions \citep{PavKirWie13}.
This produces a few grains at all distances from the ionizing star that are hotter than the equilibrium temperature and so emit more at mid-IR wavelengths.
Indeed, the model of \citet{PavKirWie13} appears to reproduce the mid-IR observations better than ours for the outer parts of the \ion{H}{ii} region (see Sect.~\ref{sec:discussion}).
\textsc{torus} does not include non-equilibrium heating of dust grains at present.

It is interesting that at long wavelengths we overpredict the emission projected within the \ion{H}{ii} region in comparison with the observations (see Fig.~\ref{fig:rcw120_160um}).
The emission arises mostly from the dense shell, projected onto the \ion{H}{ii} region because the simulations have rotational symmetry.
This suggests that in RCW\,120 there is more shell material on the projected boundaries than at the front or the rear i.e.,~it may not be an axisymmetric shell.
This has been noted before for RCW\,120 by \citet{PavKirWie13}, who suggest that perhaps RCW\,120 has an open cylindrical or ring geometry with the symmetry axis close to the line of sight.
\citet{TorHasHatEA15} argue against this on the basis of the large foreground extinction and \ion{H}{i} column density towards the region.
They argue that the \ion{H}{ii} region is on the far side of a molecular cloud and so there should be an approximately complete shell on the hemisphere facing towards us.
\citet{TorHasHatEA15} also favour a cloud-collision model in which the O star was formed in a shocked layer, and consequently it should be moving with a few km\,s$^{-1}$ with respect to the molecular cloud, as in our simulation WV04.
It is not clear how these two models for the region can be reconciled.

\begin{figure*}
\centering
\includegraphics[width=0.45\hsize]{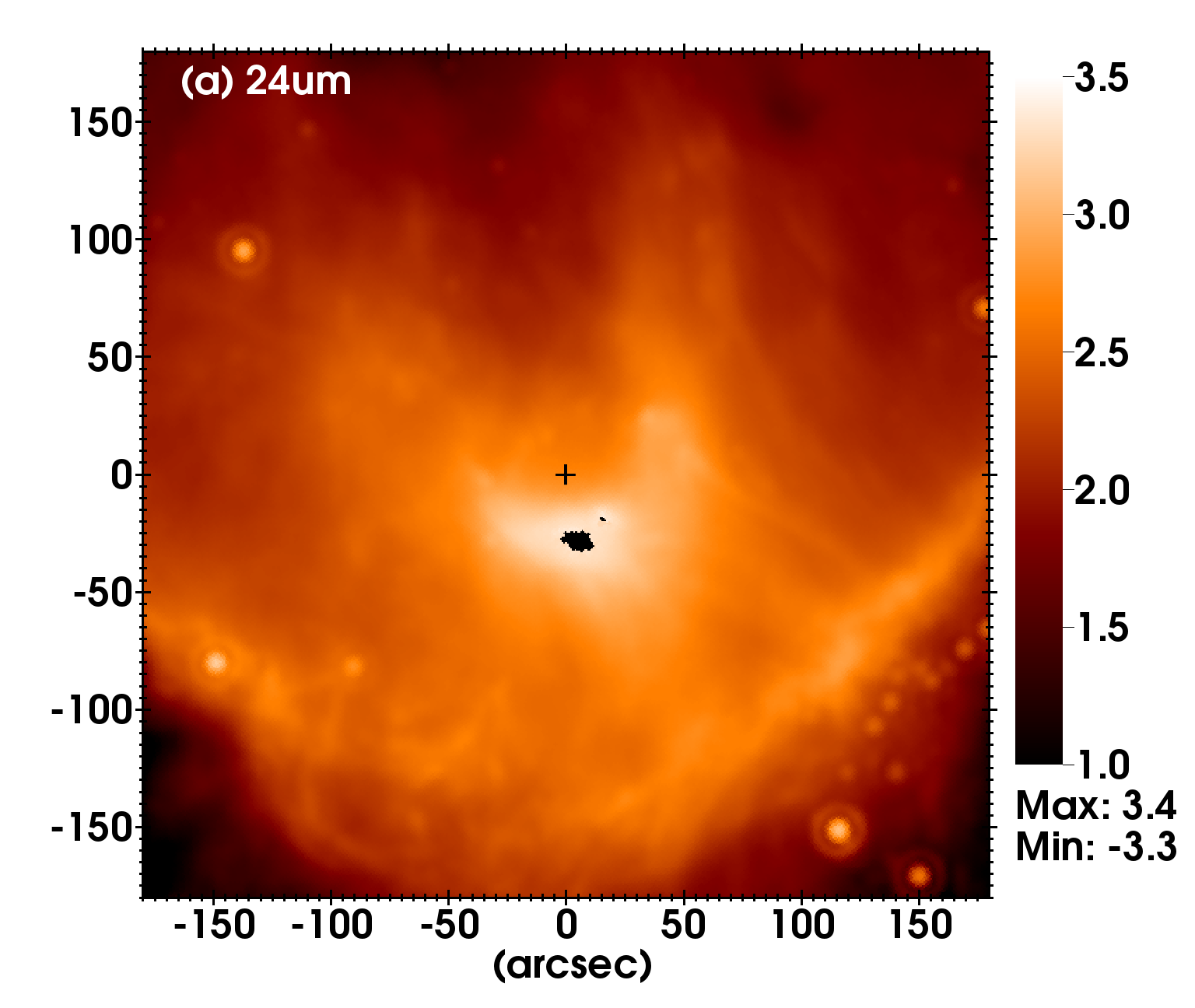}
\includegraphics[width=0.45\hsize]{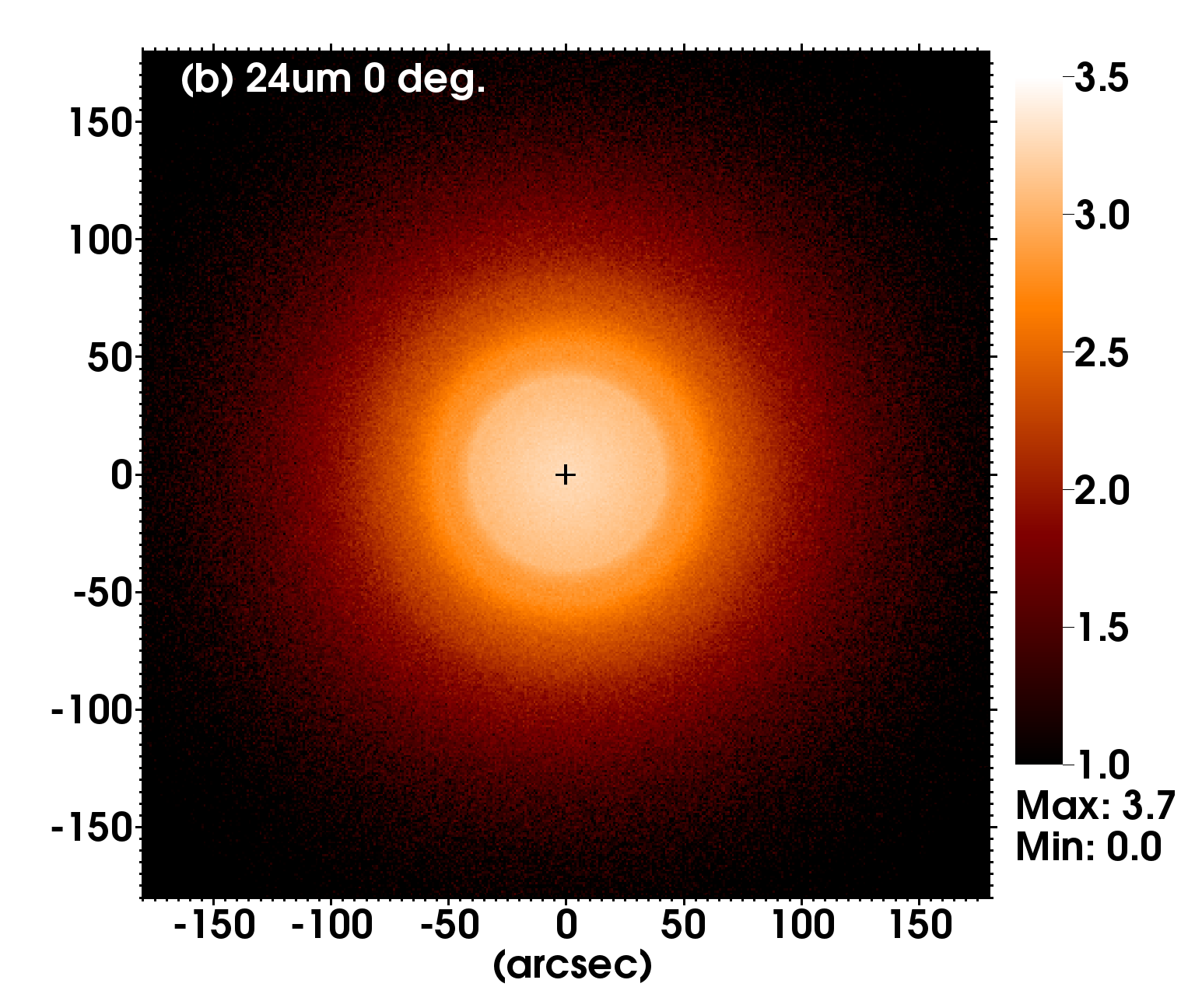}
\includegraphics[width=0.45\hsize]{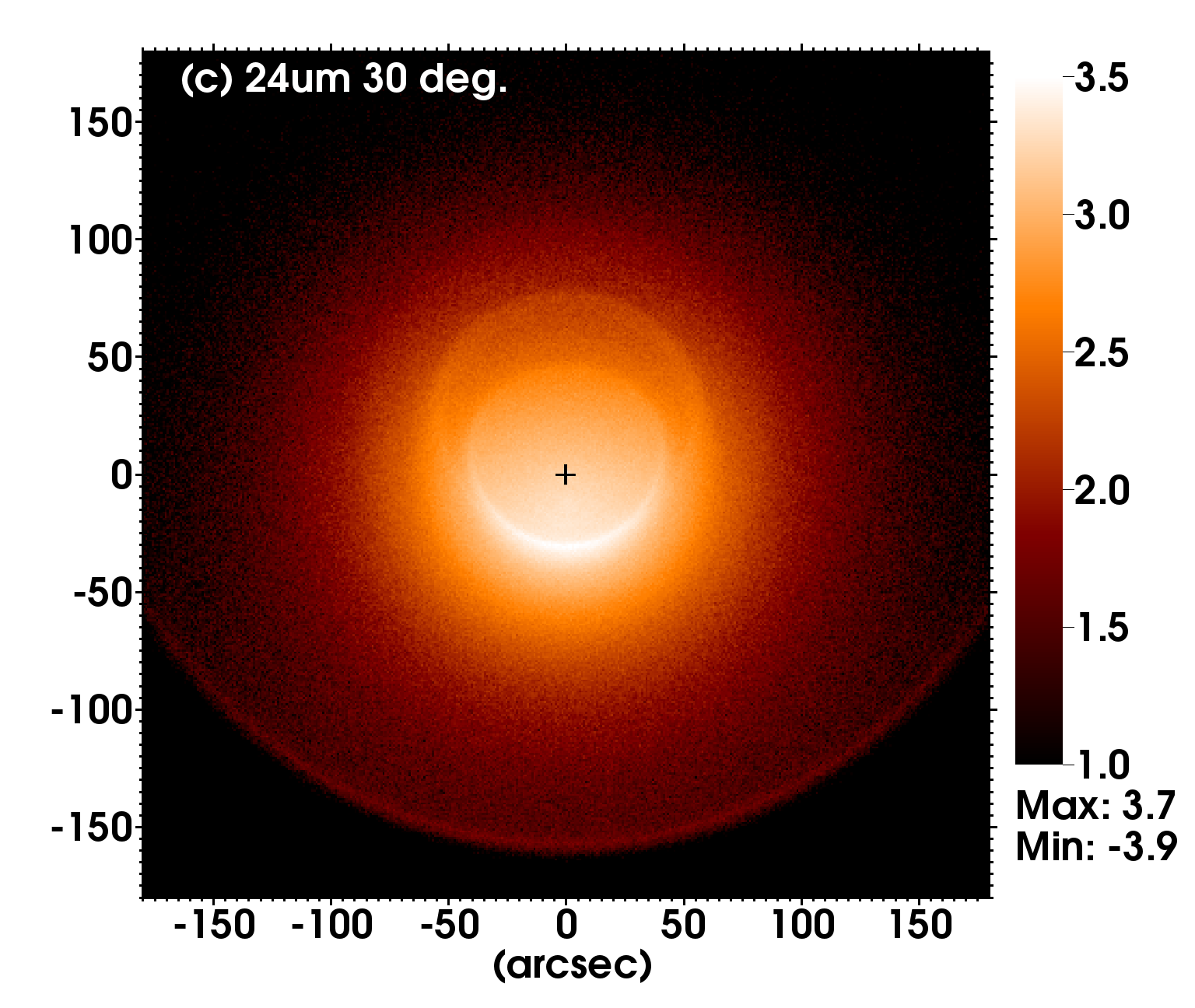}
\includegraphics[width=0.45\hsize]{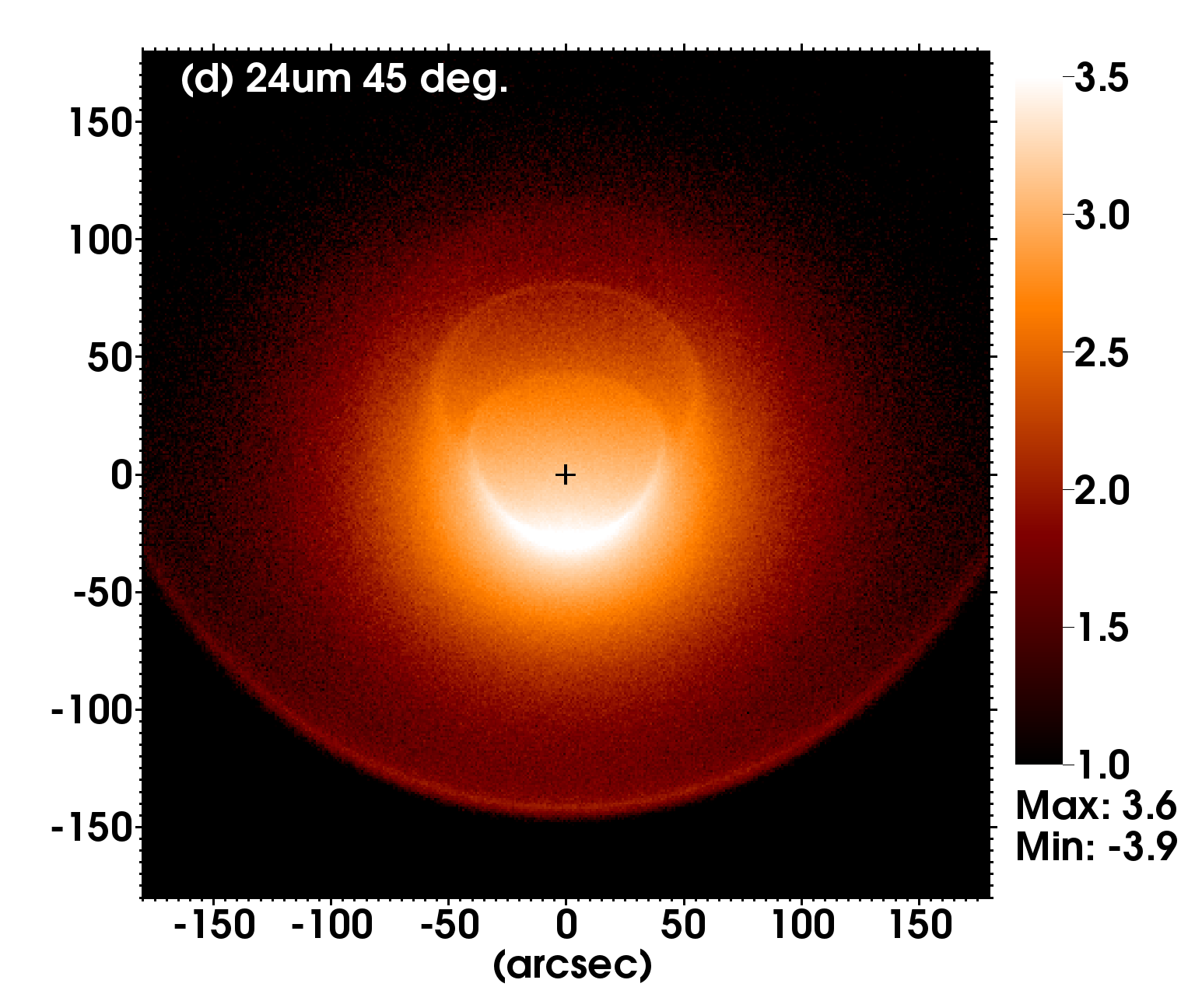}
\includegraphics[width=0.45\hsize]{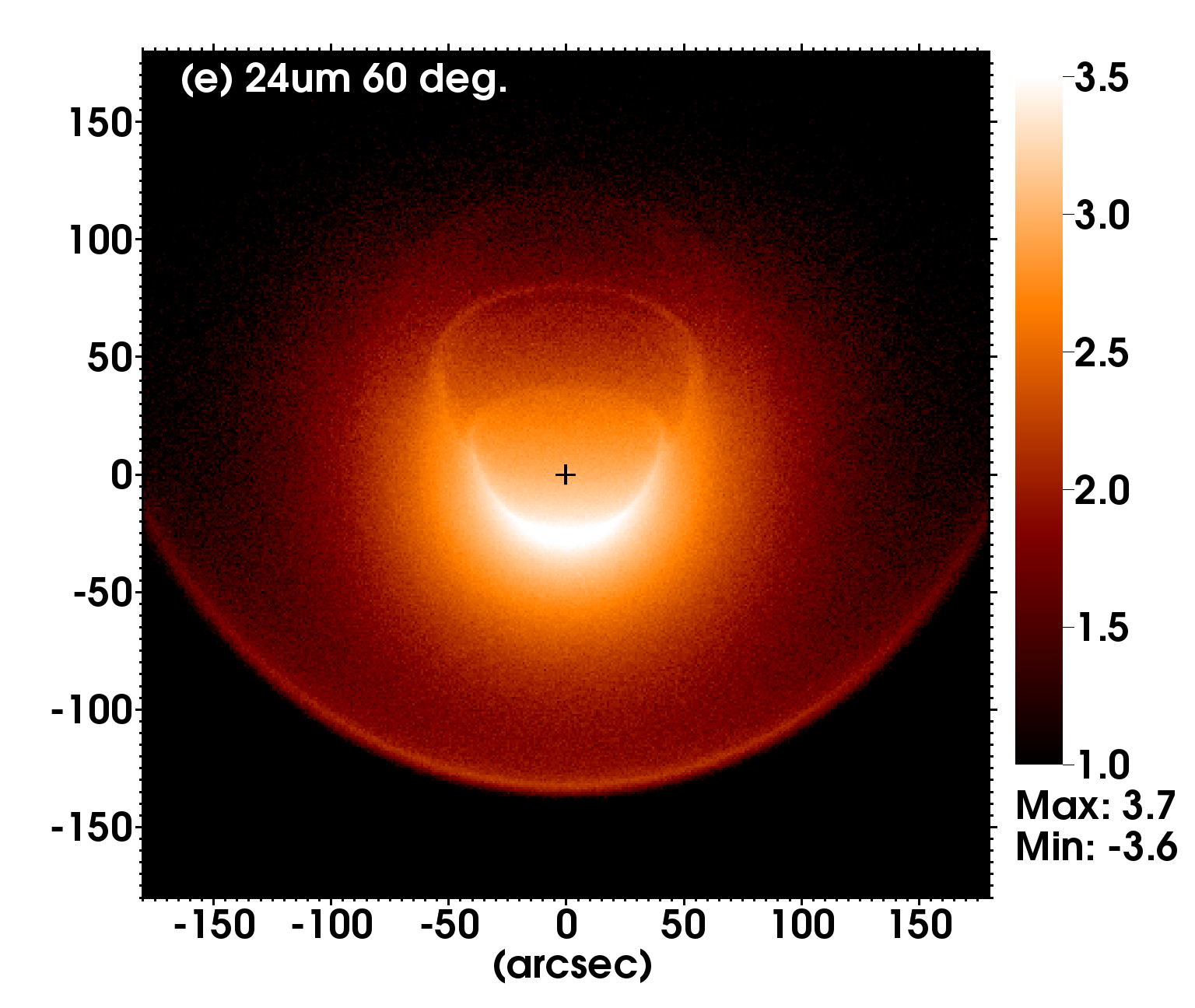}
\includegraphics[width=0.45\hsize]{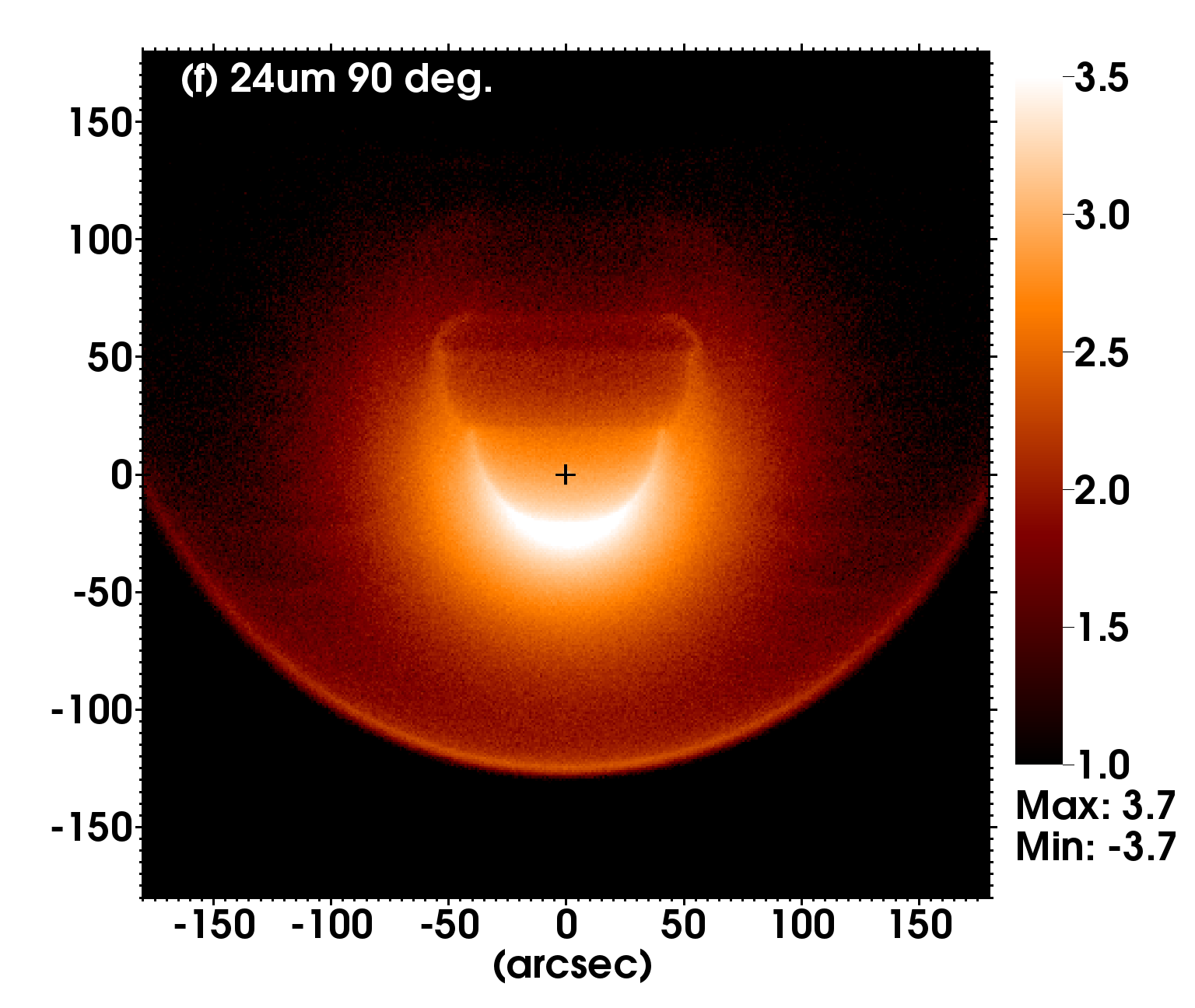}
\caption{
  \textbf{(a)} The Galactic \ion{H}{ii} region RCW\,120 in the mid-IR from \emph{Herschel} at 24 $\mu$m.
  \textbf{(b-f)} Synthetic observation at 24 $\mu$m of dust emission from the  simulation WV04 (sil2.0), at angles \textbf{(b)} 0$^\circ$, \textbf{(c)} 30$^\circ$, \textbf{(d)} 45$^\circ$, \textbf{(e)} 60$^\circ$ and \textbf{(f)} 90$^\circ$.
  The coordinates and units are as in Fig.~\ref{fig:rcw120_24um}.
  }
\label{fig:rcw120_angles}
\end{figure*}

Fig.~\ref{fig:rcw120_angles} shows synthetic 24 $\mu$m images at different orientations to the line of sight, in comparison with the observed nebula.
The comparison favours angles, $\theta$, of the symmetry axis to the line of sight with $\theta\geq45^\circ$.
When seen along the symmetry axis of the simulation (pole-on), the emission is circularly symmetric and indistinguishable from a circular bubble.
The asymmetry is clearly visible already at $\theta=30^\circ$, and becomes more extreme as $\theta\rightarrow90^\circ$, similar to the synthetic emission measure maps of bow shocks modelled by \citet{ArtHoa06}.

\begin{figure}
\centering
\includegraphics[width=0.85\hsize]{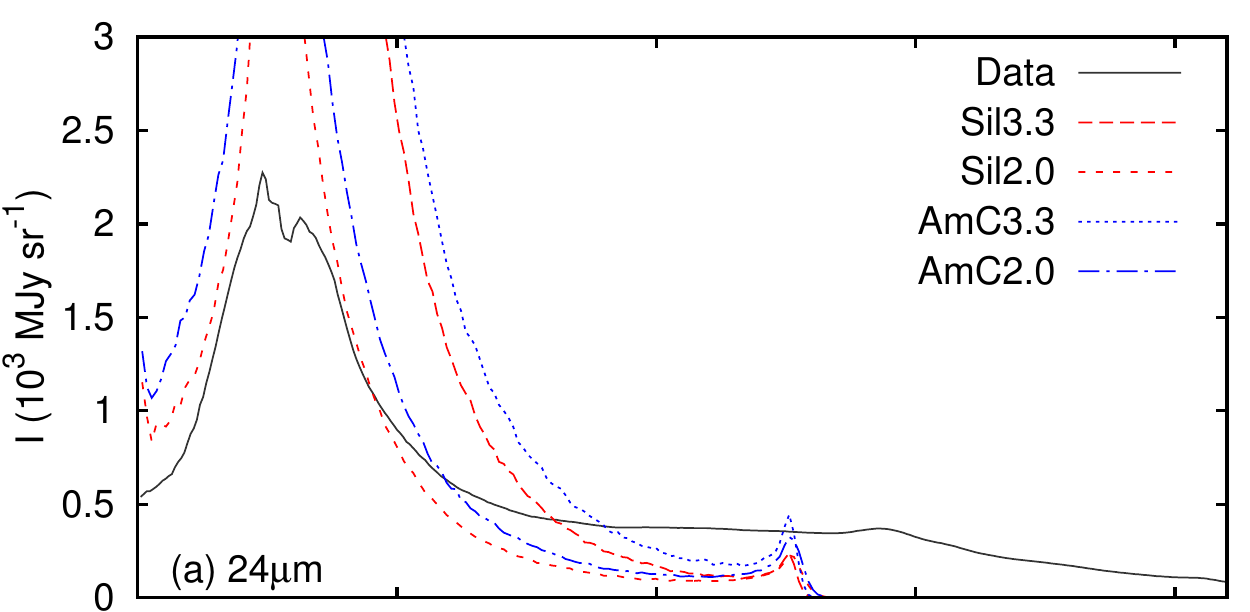}
\includegraphics[width=0.85\hsize]{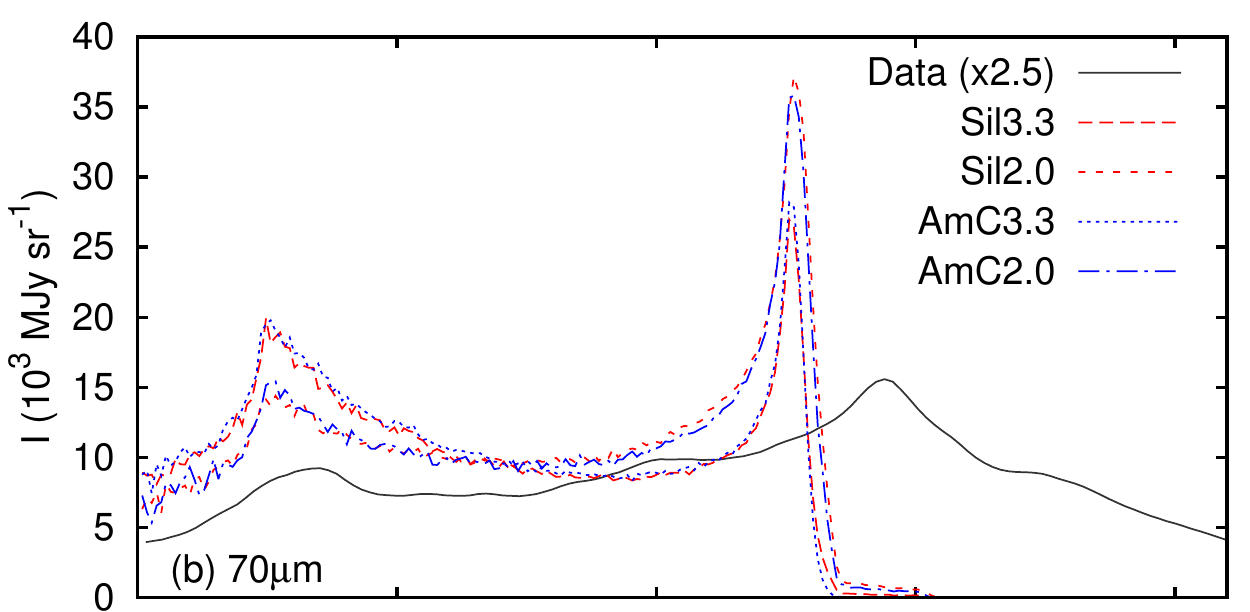}
\includegraphics[width=0.85\hsize]{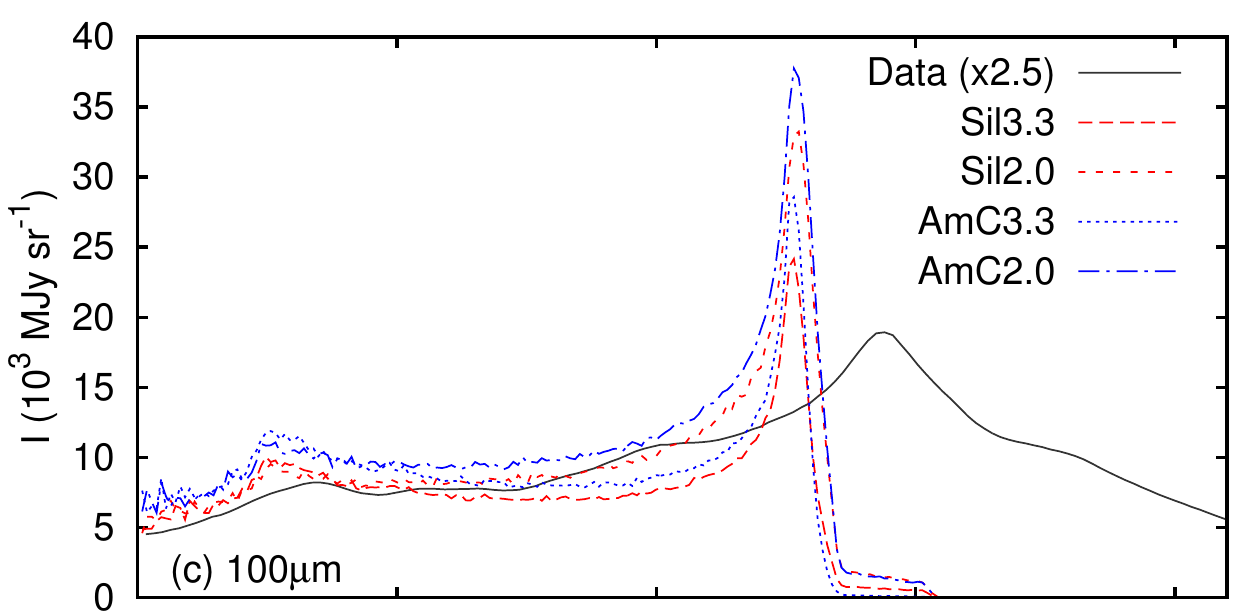}
\includegraphics[width=0.85\hsize]{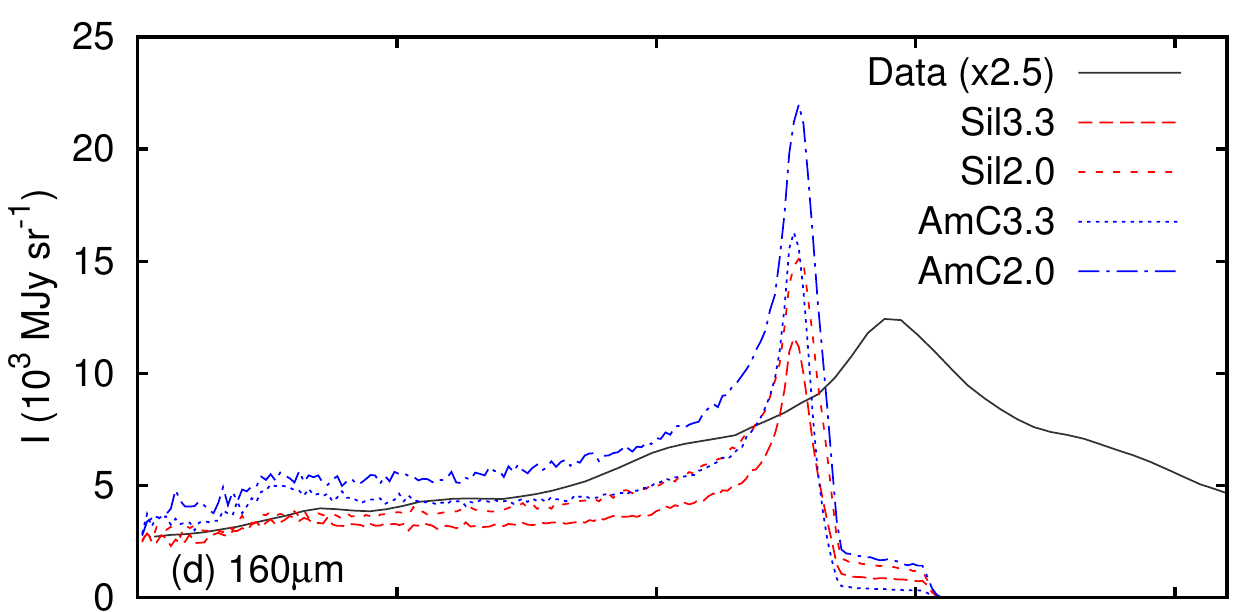}
\includegraphics[width=0.85\hsize]{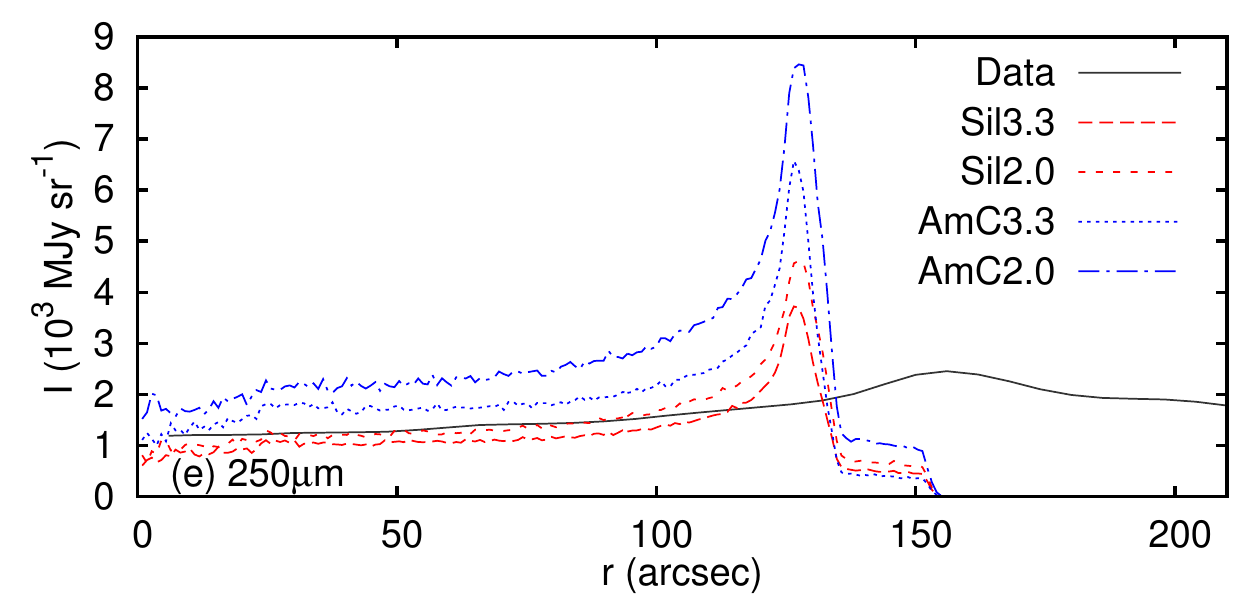}
\caption{
  Comparison of dust emission for an angle-averaged 20$^\circ$ wide wedge from the star, showing log of the intensity (in MJy\,sr$^{-1}$) as a function of angular distance, $r$, from the ionizing star (in arcseconds).
  Panels show (a) 24 $\mu$m emission, (b) 70 $\mu$m, (c) 100 $\mu$m, (d) 160 $\mu$m, and (e) 250  $\mu$m.
  Data for RCW\,120 are shown as the solid black line (multiplied by 2.5 in panels (b), (c), and (d) for clarity), with synthetic emission maps for different grain types and size distributions as the red and blue lines.
  }
\label{fig:dustV04}
\end{figure}

\begin{figure}
\centering
\includegraphics[width=\hsize]{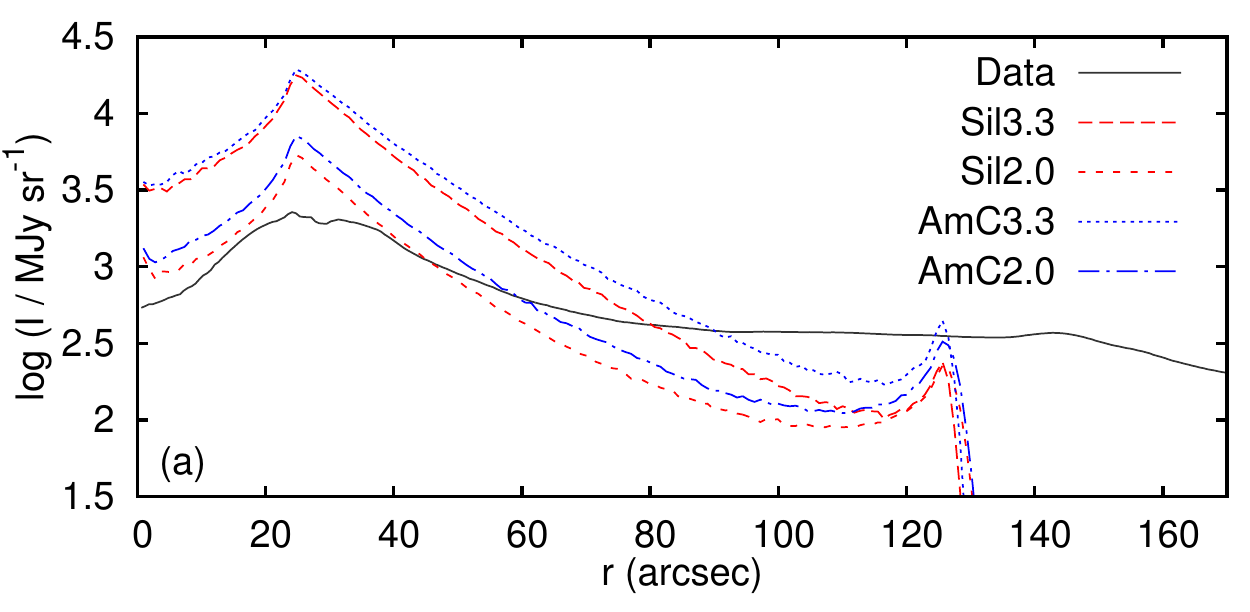}
\includegraphics[width=\hsize]{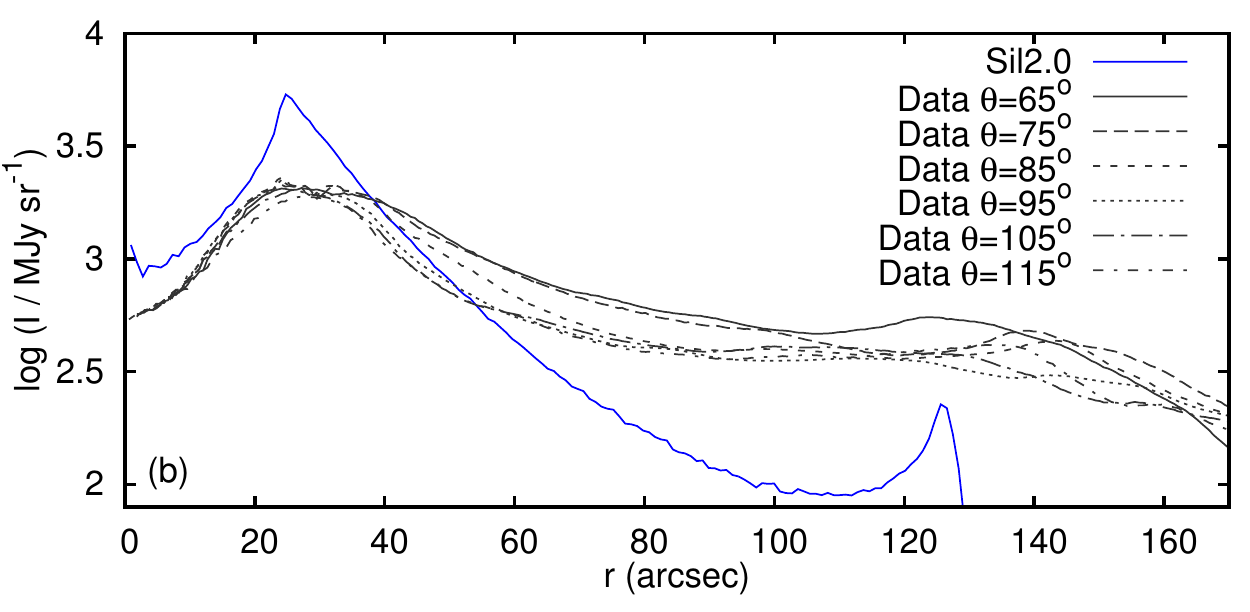}
\caption{
  (a) As Fig.~\ref{fig:dustV04}(a) but showing 24 $\mu$m emission on a logarithmic intensity scale.
  (b) Comparison of the sil2.0 synthetic map with narrow wedges of emission from different parts of the RCW\,120 \ion{H}{ii} region.
  The angles are measured from West in a clockwise direction.
  }
\label{fig:dustV04_log}
\end{figure}

\subsection{Averaged emission as a function of distance from the star}

We take the synthetic dust emission images from WV04 at $t=0.4$ Myr, using dust models Sil3.3, Sil2.0, AmC3.3, and AmC2.0 as a representative sample of the full list in Table~\ref{tab:dustmodels}.
For comparison with observations we consider emission in ``pie slices'' of the \textsc{fits} images, outwards from the star's location with an opening angle $\theta$ (in degrees, full width).
From these we obtained an average brightness as a function of distance from the star.
The data have a lot more small-scale structure than the simulations, and we wish to average over this.
Results for $\theta=20^\circ$ are presented; for the observations the wedge is centred on South (downwards in Fig.~\ref{fig:rcw120_24um}), and for the simulations it is centred on the axis of symmetry (again downwards in Fig.~\ref{fig:rcw120_24um}).
Fig.~(\ref{fig:dustV04}) shows the results for (a) 24 $\mu$m, (b) 70 $\mu$m, (c) 100 $\mu$m, (d) 160 $\mu$m, and (e) 250 $\mu$m, respectively.
We can see that there are some similarities and differences when comparing the simulations and the data, already noted from the 2D images, which we discuss in more detail below.

\subsubsection{\ion{H}{ii} region shell}
The simulations have the outer edge of the stellar wind bubble at $r\approx25$ arcsec and the dense shell surrounding the \ion{H}{ii} region at $r\approx125$ arcsec.
The observed dense shell is at $r\approx140-150$ arcsec, but in contrast to the simulations, it is quite broad and broken.
This is presumably because of ionization front instabilities and/or density structure in the molecular gas from e.g.,\ turbulence.
The offset between simulated and observed shells, and the narrower peak of the simulated shell emission, can therefore be understood simply as limitations of the hydrodynamic simulations.
For the offset, we may have used a mean density that is somewhat too large for the ISM, resulting in an \ion{H}{ii} region that is a bit too small (see discussion in Paper I).
Regarding the clumpiness of the shell, the simulations deliberately suppress instability in the shell to prevent symmetry-axis problems, and also consider a uniform ISM with no turbulence (because one cannot sensibly drive turbulence in axisymmetry).
For these reasons, we should not expect close agreement between simulations and observations for the shell; 3D simulations including turbulent pressure \citep[cf.][]{ArtHenMelEA11, DalNgoErcEA14} and/or substructure \citep{WalWhiBisEA12, WalWhiBisEA13, WalWhiBisEA15} in the ISM would be required.

\subsubsection{Inner arc of 24 $\mu$m emission}
Similar to WV00, simulation WV04 shows two emission peaks at short wavelengths (24 and 70 $\mu$m), corresponding in the synthetic emission maps to the outer edge of the wind bubble and to the \ion{H}{ii} region shell.
At longer wavelengths, the edge of the wind bubble is no longer apparent, and at all radii we are seeing mainly emission from the dense shell projected on top of weak emission from the \ion{H}{ii} region interior.
Fig.~\ref{fig:dustV04}(a) shows 24 $\mu$m emission as a function of distance from the ionizing star on a linear scale, and the same data are plotted on a logarithmic scale in Fig.~\ref{fig:dustV04_log}.
We predict too much emission near the star by an order of magnitude at 24 $\mu$m for dust models Sil3.3 and AmC3.3 (although MIPS has some saturated pixels here, so may be missing some flux), whereas Sil2.0 shows very good agreement with the data for $r<50$ arcsec.
This is simply a normalisation issue, where the gas-to-dust ratio would come into play along with the dust temperature and composition, so we should not expect to obtain better agreement than what we see here.

At larger radii ($r>50$ arcsec) the simulated emissivity decreases with distance much more dramatically than the data.
The simulated emission decreases exponentially with increasing distance, as for WV00 (Fig.~\ref{fig:compV00}), out to radii where contributions from the dense shell become important ($r\gtrsim100$ arcsec), at which point the emission levels off.
There is then a small peak at the inner edge of the dense shell; this is a density effect.
The data, however, show a more gentle decrease from the peak at $r\approx30$ arcsec to an approximately constant value for $r\gtrsim80$ arcsec, followed by an exponential dropoff outside the dense shell at $r>150$ arcsec.
This exponential decrease at $r>150$ arcsec is seen in all panels of Fig.~\ref{fig:dustV04} except panel (e) (250 $\mu$m) and presumably reflects the decreasing dust temperature further from the ionizing star of RCW\,120.

\subsubsection{Emission at intermediate wavelengths, 70, 100, and 160 $\mu$m}
At the three intermediate wavelengths, 70, 100, and 160 $\mu$m (panels b, c, and d of Fig.~\ref{fig:dustV04}, respectively) we predict too much dust emission.
The observational data are multiplied by a factor of 2.5 so that we can show them on the same scale.
At these wavelengths the brightness is less sensitive to temperature and more to the column density of dust.
A change in the dust-to-gas ratio or a moderate change in overall mean gas density would bring the data and synthetic observations into very good agreement (apart from the aforementioned discrepency with the outer shell).
We note that the Sil2.0 dust model provides a better fit than Sil3.3 or AmC3.3 at these wavelengths, as well as at 24 $\mu$m.
This is because the peak at 30 arcsec is less pronounced and the increase in flux with $r$ in $50\lesssim r\lesssim125$ arcsec is much better reproduced.
The disagreement at 160 $\mu$m is likely because the observed \ion{H}{ii} region shell is much broader than in the simulations.

\subsubsection{Emission at 250 $\mu$m}
At 250 $\mu$m the normalisation of the emission comes more into agreement with the simulations, at least for silicate grains.
The synthetic data reproduce the approximately flat emission profile seen in the observations, failing only at the \ion{H}{ii} shell.
The simulations have a thin and dense shell that has significant dust emission, and which absorbs most of the radiation from the ionizing star.
This means that there is very little emission predicted for gas outside the shell.
RCW\,120 has a broken and porous shell, in contrast to our simulation, so more FUV radiation escapes into the lower-density ISM and therefore provides more dust heating.
This is obviously not a problem with the dust model, but rather a difference between the simulated and observed shell structures.
The difference may also partly arise because we only consider one source of radiation, and do not consider heating by the interstellar radiation field and other young low-mass stars in the region, which probably becomes important outside the shell.

\section{Discussion}
\label{sec:discussion}

In Paper I we highlighted that there are multiple explanations for the IR arcs, which we here refer to as \textit{displacement}, \textit{mixing}, and \textit{decoupling} models, described in Sect.~\ref{sec:intro}.
Our work explores the displacement model and so we discuss this in more detail below, comparing and contrasting our results with previous work, and also discussing our work in the context of the other models.

\subsection{Displacement model for dust in \ion{H}{ii} regions}

\citet{WatPovChuEA08} suggested that the ring of 24 $\mu$m within the \ion{H}{ii} region N49 shows a hole in the dust emission that has been evacuated by a SWB.
This corresponds to a \textit{displacement} model, where wind and ISM do not significantly mix
except at a turbulent and thermally conducting boundary layer, and mid-IR emission arises from the \ion{H}{ii} region \textit{outside} the SWB.
In Paper I we used RHD simulations to show that a stellar wind cavity is also a plausible explanation for the IR arc in RCW\,120 (see Fig.~\ref{fig:rcw120_24um}), for a wind strength predicted by theory \citep{VinDeKLam01} and compatible with observational upper limits \citep{MarPomDehEA10}, and for an ISM number density $n_\mathrm{H}=3000$ cm$^{-3}$.
This was based only on the size and shape of the wind bubble that is produced, not on any predicted emission properties.
Here we have confirmed this result, showing that the mid-IR arc of emission in RCW\,120 is morphologically very similar to that obtained in synthetic emission maps produced by our simulations.
We find that the outer edges of SWBs are probably most easily detected by their IR emission, agreeing with the study of bow shocks by \citet{MeyMacLanEA14}, who found that all of their simulated bow shocks emitted more in the IR than in any other tracer.

As noted in Sect.~\ref{sec:synthetic_images}, \citet{TorHasHatEA15} propose a model in which CD\,$-$38$\degr$11636 should move at a few km\,s$^{-1}$ relative to the molecular cloud.
Using the most recent proper motion measurements for CD\,$-$38$\degr$11636 provided by the PPMXL \citep{RoeDemSch10} and SPM 4.0 \citep{GirVanZacEA11} catalogues, we derived components of the peculiar transverse velocity of the star (in Galactic coordinates) of $v_\mathrm{l}=-29.9\pm17.3$, $v_\mathrm{b}=39.4\pm17.3$ km\,s$^{-1}$ and $v_\mathrm{l}=1.6\pm16.6$, $v_\mathrm{b}=14.1\pm16.6$ km\,s$^{-1}$, respectively.
To derive these velocities, we used the Galactic constants from \citet{ReiMenZheEA09a}
and the solar peculiar motion from \citet{SchBinDeh10}.
For the error calculation, only the errors of the proper motion measurements were considered.
The obtained velocities are consistent with our inference that the space velocity of CD\,$-$38$\degr$11636 is only several km\,s$^{-1}$ at the 2 and 1 sigma level, respectively.

\citet{PavKirWie13} used spherically symmetric simulations of expanding \ion{H}{ii} regions without a SWB, together with radiative transfer post-processing similar to what we have done, to compare synthetic emission maps with observations.
They make similar assumptions to ours: dust and gas are dynamically coupled, and the grains are not destroyed within the \ion{H}{ii} region (except for PAH, which we do not model and which do not contribute to the mid- and far-IR emission presented here).
They additionally tested whether collisional heating of the dust by the gas is significant, finding that it was negligible compared to radiative heating.
Their results are similar to our spherically symmetric results HV00, and they noted that a wind bubble (or some other process) is required to create the inner arc of emission in RCW\,120.

The model of \citet{PavKirWie13} includes stochastic heating of small dust grains, and their fig.~(4) shows that some very small grains can have temperatures $>100$ K even 1.5 pc from the ionizing star, where the equilibrium temperature is $\approx20-30$ K.
This probably explains why they obtain more 24 $\mu$m emission far from the star than we do: $\sim500$ MJy\,sr$^{-1}$ compared with our $\sim200$ MJy\,sr$^{-1}$.
This is also in better agreement with observations; we noted that Fig.~\ref{fig:dustV04_log} shows a deficit of predicted 24 $\mu$m emission compared with observations for positions $\gtrsim70$ arcseconds from the ionizing star, whereas \citet{PavKirWie13} find better agreement.
This suggests that our \textsc{torus} results might come into even better quantitative agreement with observations if we included stochastic grain heating.

At long wavelengths, \citet{PavKirWie13} find a thinner \ion{H}{ii} region shell than is observed, probably for the same reasons as in our models: the observed shell is broken up by instabilities and/or pre-existing inhomogeneity.
Although they did not calculate mid- or far-IR dust emission maps, the simulations by \citet{ArtHenMelEA11} of \ion{H}{ii} region expansion in a turbulent medium show a broken shell that is broadened when seen in projection.
Simulations of \ion{H}{ii} regions expanding in a fractal medium by \citet{WalWhiBisEA15} show a similar picture when seen in projection and in synthetic images at 870 $\mu$m.
While this is unlikely to affect our predictions for the inner arc of mid-IR emission, such inhomogeneities are an essential component of a full model for the wind bubble and \ion{H}{ii} region around a young star in a dense environment.

\citet{ArtHoa06} and \citet{ZhuZhuLiEA15} have studied the dynamics of bow shocks and \ion{H}{ii} regions around moving stars, comparing the results with what they find for static stars embedded in a density gradient.
In both simulations an asymmetric SWB is produced, although there are differences in the dynamics of the ionized and neutral gas.
It is not obvious that there would be differences in the dust emission produced, and we expect in both cases that a bright arc of 24 $\mu$m emission would be produced where the distance of the SWB edge from the star is a minimum.
This should be explored in future work.

\subsection{Mixing model for dust in \ion{H}{ii} regions}

For the N49 \ion{H}{ii} region, \citet{EveChu10} proposed a different model to interpret the observations of dust within the \ion{H}{ii} region.
In their model the \ion{H}{ii} region is almost entirely filled with shocked stellar wind that is polluted (mass-loaded) by photoevaporating ISM globules, continuously adding dust to the hot gas.
This is a \textit{mixing} model, originally proposed by \citet{McKVanLaz84}, where the dusty and clumpy ISM is efficiently mixed with shocked stellar wind and heated by collisional and radiative processes \citep[see also][]{ArtDysHar93, WilDysRed95}.
In this model, the hole at the centre arises because the wind has displaced all ISM material in the 0.5$-$1 Myr age of the bubble \citep{EveChu10}.
It thus incorporates the displacement model, but additionally has a very broad mixing region that takes up most of the volume of the \ion{H}{ii} region, and the mid-IR emission arises from this mixing region.

Mass-loading of O star winds by evaporation of protostellar disks could also inject significant quantities of gas and dust into the SWB of RCW\,120, as in the Orion Nebula \citep{Art12}.
If this were the case, we would expect to see mid-IR emission that increases in brightness towards CD\,$-$38$\degr$11636, instead of the arc of emission that is observed.
This implies that either there are no protostars within the SWB, or that the evaporated dust is destroyed very effectively in a region around the star \citep{EveChu10}.
If we apply the model of \citet{EveChu10} to RCW\,120, the inner arc would be interpreted as marking the interface between the free-wind region and the hot bubble (because the latter is full of photoevaporated dust).
The difficulty with this interpretation is that there remains no other observed feature in dust emission that could represent the contact discontinuity, although it should be visible because of the huge increase in density.
This could be further investigated using multi-dimensional simulations of mass-loaded winds following \citet{Art12}, but including a model for the destruction of dust grains.

\citet{EveChu10} also demonstrated that dust sublimation is not an issue in \ion{H}{ii} regions, and showed that sputtering timescales are 0.01-1\,Myr for grains in a hot bubble with $T=3.5\times10^6$\,K.
For grains in the \ion{H}{ii} region, with $T\approx10^4$\,K, the sputtering times are orders of magnitude longer \citep[][fig.~12]{TieMcKSeaEA94}.
These results show that dust destruction processes are not expected to have a significant effect on our results within the displacement model.

We note that simulations using a uniform density ISM, such as those in Paper I and this work, automatically favour the displacement model because of the smooth initial conditions.
By contrast, models with very clumpy or turbulent initial conditions such as those of \citet{RogPit13} or \citet{DalNgoErcEA13} favour the mixing model, especially when ionizing radiation is not considered.
Photoionization tends to homogenise the density distribution by heating everything within the \ion{H}{ii} region to $T\approx10^4$ K, although how effective this is depends somewhat on the properties of the ionizing star \citep{ArtHenMelEA11, DalNgoErcEA14}.
Further work is required to determine whether ISM turbulence and substructure \citep{MedArtHenEA14} has a significant effect on SWBs that are contained within \ion{H}{ii} regions.
As an example, NGC\,7635 has an apparently spherical SWB even though is it located right at the edge of a molecular cloud with many pillars and bright-rimmed clouds nearby \citep{ChrGouMeaEA95, MooWalHesEA02}.

\subsection{Radiation pressure and the decoupling model}

Observations of dust emission in \ion{H}{ii} regions show that often there is a central cavity devoid of dust \citep{Ino02}.
\citet{KruMat09} studied radiation pressure effects, finding that they may be the dominant feedback mechanism for \ion{H}{ii} regions around massive star clusters, but not around single stars or small clusters.
\citet{Dra11} calculated equilibrium density structures for \ion{H}{ii} regions assuming that radiation pressure is balanced by gas pressure (no gas or dust motions), and showed that significant decreases in central density can arise for Galactic \ion{H}{ii} regions.
This occurs without dynamical decoupling of gas and dust and without the action of stellar winds.
In contrast, \citet{SilTen13} calculate that radiation pressure has very little effect on the gas density in expanding \ion{H}{ii} regions with stellar wind bubbles, implying that the \citet{Dra11} density structure may not have time to become established in real \ion{H}{ii} regions.
\citet{PalUmaVenEA12} presented IR observations of many \ion{H}{ii} regions, showing that the warm dust emitting at 24 $\mu$m is difficult to explain in the context of the \citet{Dra11} predictions.
They argue in favour of replenishment of dust in hot gas, photoevaporated from dense neutral clumps, as advocated by \citet{EveChu10}.
Replenishment is not required if radiation pressure is not effective.

\citet{OchVerCoxEA14} propose that the IR arc in RCW\,120 has nothing to do with a SWB, but rather that radiation pressure from the ionizing star on dust grains excludes dust from a region near the star.
This creates a \textit{dust wave}, where the dust density (and hence IR emission) increases even though the gas density is unchanged, because the dust is dynamically decoupled from the gas.
This is therefore a \textit{decoupling} model, and these authors argue that it could explain many IR arcs and rings in \ion{H}{ii} regions.
They did not make a quantitative comparison with the inner arc of RCW\,120, however.
Furthermore, they assume that the \ion{H}{ii} region is 2.5 Myr old, whereas in paper I we presented strong arguments (in agreement with most previous work) that RCW\,120 is a young \ion{H}{ii} region with age $<0.5$ Myr \citep[cf.][]{ArtHenMelEA11}.

\citet{OchCoxKriEA14} and \citet{OchTie15} also apply their model to a mid-IR arc around $\sigma$ Ori, finding that the decoupling model provides a good match to observations because there is no corresponding structure seen in gas emission.
We have shown that one can still obtain a bright arc of dust emission even when there is no corresponding increase in gas density (hence emission measure), by allowing a SWB to exist.
\citet{OchTie15} found that there are in fact two arcs of dust emission around $\sigma$ Ori, and interpret this as evidence of two different populations of dust forming two distinct dust waves.
It is also possible, however, that one arc represents the edge of an asymmetric SWB ($\sigma$ Ori is embedded in a Champagne flow) and the other is a dust wave.
If the Champagne flow is subsonic, then there would be no bow shock, but we have shown that there would be a mid-IR arc of dust emission from a stellar wind.

Estimating the degree of coupling between dust and gas is in itself an active field of research \citep{Dra11,VanMelKepEA11, PalUmaVenEA12, OchCoxKriEA14, AkiKirPavEA15, PriLai15} and has many uncertainties.
\citet{OchCoxKriEA14,OchVerCoxEA14} did not include Coulomb interactions in the gas-dust interaction; this model therefore requires that dust grains are efficiently decharged in \ion{H}{ii} regions.
\citet{AkiKirPavEA15} have investigated the coupling of gas and dust in \ion{H}{ii} regions in a more general framework, calculating the charge distribution of dust grains, and the differences between models where dust is charged and uncharged.
They find that differential grain charge as a function of distance from the centre of an \ion{H}{ii} region has a strong effect on the gas-dust coupling, and that small grains (which may produce most of the mid-IR emission) are effectively coupled to the gas.
Application of the \citet{AkiKirPavEA15} calculations to the Champagne flow model of \citet{OchVerCoxEA14} would be very useful, to determine how robust the dust-wave predictions are to the effects of grain charge.
None of these studies include the Lorentz force which couples charged grains to the local magnetic field, and it is unclear how much this would change their predictions.
At present we see no clear reason to favour either the displacement or decoupling model for the inner arc in RCW\,120 or around $\sigma$ Ori.
Future work should focus on investigating both models more completely to find ways of distinguishing between them.

\subsection{Why are so few wind bubbles detected in tracers of gas emission?}

Most (if not all) massive stars form in a clustered way \citep{LadLad03}, and so their winds expand into cavities created by the winds of their companion stars; individual bubbles are unlikely to be observed in such an environment unless a star is embedded in a supersonic gas flow driven by the cluster, in which case a bow shock is sometimes seen \citep{AscAlvVicEA07, PovBenWhiEA08, KobGilKim10}.
On the other hand, about 20\% of Galactic O and B stars are located outside of clusters and OB associations \citep{Gie87}.
These stars are isolated through dissolution of their parent clusters or ejection from their birth places either by dynamical few-body encounters \citep{PovRuiAll67, GieBol86, OhKroPfl15} or binary supernova explosions \citep{Bla61, Sto91, EldLanTou11}.

Stars that only recently escaped from embedded star clusters move supersonically through the dense and cold gas of their parent molecular clouds and form bow shock-confined, cometary, ultracompact \ion{H}{ii} regions \citep{MacVanWooEA91, ArtHoa06, ZhuZhuLiEA15}, which are visible only at radio wavelengths \citep{WooChu89}.
After crossing the cloud, their SWBs expand into the lower density ISM.
The ISM at the bubble edge is shocked by the bubble expansion, and so the shocked ISM is denser, and hence brighter, than the undisturbed ISM.
According to the argument of \citet{DysdeV72}, the newly formed SWBs can be easily detected only while they are young and their expansion is supersonic.

In the low-density ISM, isolated O stars tend to be exiles from clusters, moving rapidly through the ISM and so their asymmetric SWBs transform into bow shocks \citep{GulSof79, VanMcC88,  KapVanAugEA97, GvaBom08, MeyMacLanEA14}.
Paper I showed that even low-velocity O stars quickly produce asymmetric wind bubbles even for subsonic motion where there is no bow shock.
These two factors
(that bubbles are only spherical \textit{and} bright when they are very young)
are responsible for the paucity of circular bubbles around isolated massive main sequence stars.
Only one such bubble is known in the Galaxy: NGC\,7635 or the Bubble Nebula \citep{ChrGouMeaEA95, MooWalHesEA02}.

Large SWBs can be detected in gas emission as ``holes'': lines of sight with lower emission measure than their surroundings because of the very low density in the bubble.
This is seen, for example, in the N49 bubble in 20 cm radio emission \citep{WatPovChuEA08}, and on a larger scale in the Rosette Nebula (\citealt{Mat66} and references therein; see also \citealt{SavSpaFis13}).
For this to be possible, however, the cavity should be about half the diameter of the larger \ion{H}{ii} region.
If it is much smaller, then lines of sight through the cavity have similar emission measure to those outside the cavity, whereas if it is much larger then less and less of the \ion{H}{ii} region has lines of sight that do not pass through the cavity.
For our simulations WV04 and HV04, emission measure maps do not show any significant differences near the IR arc, showing that such maps cannot refute the presence of a SWB.

Faraday rotation can increase sensitivity to cavities because it traces the magnetic field as well as the electron density.
If the cavity has a weak magnetic field (expected unless field lines can cross the contact discontinuity from the ISM into the cavity) and has displaced the pre-existing ISM field, then the rotation measure can show quite clearly the edge of a SWB \citep{IgnPin13, SavSpaFis13, PurGaeSunEA2015}.
This is a promising tool for detecting SWBs in cases where the contrast in emission measure is not sufficient to detect a low-density cavity.

SWBs can also be detected as holes in \ion{H}{i} emission \citep[e.g.,][]{CapBen98, CapHer00}, but this only works if the wind bubble fills the \ion{H}{ii} region completely; otherwise the \ion{H}{i} hole traces the \ion{H}{ii} region.
Late-O stars such as CD\,$-$38$\degr$11636 in RCW\,120 have winds that are too weak to fill their \ion{H}{ii} region when on the main sequence \citep[][]{WeaMcCCasEA77}.

\subsection{The wind-ISM boundary}

The contact discontinuity between stellar wind and ISM can have a density jump of a factor $10^3-10^4$ because of the temperature difference.
This is similar to the density contrast between air and water, or even rock, and suggests naively that mixing at the boundary layer may be limited.
It is clear, however, that energy is efficiently transported across the SWB boundary based on observations \citep{RosLopKruEA14}, possibly by thermal conduction \citep{WeaMcCCasEA77} or turbulent mixing (Paper I).
Mixing and/or diffusivity is certainly important, because simulations show that the size of the wind bubble depends quite strongly on how the contact discontinuity is treated in terms of numerical diffusivity \citep{MacMcCNor89, VanKep11} and whether thermal conduction is important or not \citep{ComKap98, MeyMacLanEA14}.
\citet{MeyMacLanEA14} showed that the volume occupied by stellar wind material is decreased significantly when thermal conduction is included (i.e.~when the physical diffusivity is increased), because the thermal pressure that maintains the wind bubble is transported to the ISM.
\citet{ZheMya98} showed that conduction also significantly alters the structure of spherically symmetric SWBs around static stars.
Spectral signatures of a conduction front at the edge of the SWB around the Wolf-Rayet star HD\,50896 were detected by \citet{BorMcCClaEA97}, although further work is required to constrain the strength of the conduction, which can be affected by e.g., magnetic fields \citep{BorBalFri90}.
Our simulations do not include thermal conduction, so the mixing layer is mediated by turbulent mixing, which we found to be very efficient at removing energy from the SWB.
Simulations including the effects of thermal conduction should be performed to study how this extra diffusivity could affect the edge of the SWB and its associated 24 $\mu$m arc.

\subsection{Limitations of the present model}

Interstellar magnetic fields can also affect the expansion of \ion{H}{ii} regions \citep{KruStoGar07} and stellar wind bubbles \citep{VanMelMar15}.
When the pressure of the displaced (and compressed) magnetic field becomes comparable to the driving pressure of the bubble, then the bubble becomes asymmetric.
This extra form of asymmetric pressure should also be taken into consideration in future work to constrain mass-loss rates of O stars, but is beyond the scope of this paper.

We only consider radiation from a central O star in our calculation, but in reality young \ion{H}{ii} regions also contain lower-mass protostars that emit X-rays \citep{PreKimFavEA05} because of surface magnetic activity, and EUV/FUV radiation is well-correlated with X-ray luminosity for low-mass stars \citep{LinFraAyr13}.
This introduces a diffuse heating source to the dust within the \ion{H}{ii} region and in the surrounding shell according to the distribution of protostars \citep{ZavPomDehEA07}.
It is possible that this distributed radiation field could increase the dust temperature far from the central O star and hence increase the relative brightness of the shell compared with the inner arc at 24 $\mu$m (Fig.~\ref{fig:dustV04_log}).
For example, \citet{WalWhiBisEA15} showed that it is important to consider radiative heating from embedded sources in order to estimate masses of dense clumps in the \ion{H}{ii} region shell based on far-IR emission.

Our hydrodynamic simulations do not include absorption of EUV radiation by dust (only by gas), whereas \textsc{torus} does include this.
For \ion{H}{ii} regions in dense gas, dust can absorb a significant fraction of the ionizing photons \citep[$\sim40$ per cent in dense regions][]{ArtKurFraEA04}, and so the \ion{H}{ii} region that we simulate is too large by $\approx10-15$ per cent.
In other words, \textsc{torus} attenuates more ionizing photons than \textsc{pion}, and so the outer shell at the edge of the \ion{H}{ii} region is irradiated (and photoheated) less strongly than it should be because it is too far away.
This may also play some role in the discrepency between our predicted mid-IR emission and the observed emission in RCW\,120.
In future simulations, both gas and dust absorption of EUV radiation will be included in order to make fully self-consistent dust emission maps.

\section{Conclusions}
\label{sec:conclusions}

We have run RHD simulations of wind bubbles within \ion{H}{ii} regions, for an O star in a dense medium matching to the stellar and ISM properties of the photoionized nebula RCW\,120.
We assume that dust and gas are dynamically coupled, that there is no dust creation or destruction in the simulation, that stellar wind material is dust free, and that ISM material has a constant dust-to-gas ratio.
We post-process these simulations with the Monte-Carlo, radiative-transfer code \textsc{torus} to make the first quantitative predictions for the mid- and far-IR emission from such a configuration, using different grain properties and size distributions.

Spherically symmetric calculations for a static star with no stellar wind (simulation HV00) show similar results to those obtained previously \citep{PavKirWie13}.
The main caveat is that we predict less mid-IR emission in the outer parts of the \ion{H}{ii} region, possibly because we neglect stochastic heating of small grains.
When a stellar wind is included (simulation WV00), the edges of the SWB and \ion{H}{ii} region create two IR rings of emission.
The inner ring is bright at mid-IR wavelengths and the outer one at far-IR wavelengths.
The inner ring emits brightly because of the increasing dust temperature closer to the star, whereas the outer ring is bright because it is overdense compared with its surroundings.
The quantitative details of the brightness of the two peaks at different wavelengths depends somewhat on the grain properties and size distribution, but the qualitative features of our results are robust.

Axisymmetric calculations of a star moving with 4 km\,s$^{-1}$ through the ISM show very different results because both the SWB and the \ion{H}{ii} region are now asymmetric.
The upstream part of the SWB is closest to the star and it emits very brightly in an arc of 24 $\mu$m emission that is shaped like a bow shock, even though there is no bow shock because the star's motion through its \ion{H}{ii} region is subsonic.
The SWB arc is also visible at 70 $\mu$m, but at 160 and 250 $\mu$m it can no longer be seen because the emission is dominated by the massive shell surrounding the \ion{H}{ii} region.
For parameters chosen to match the observed properties of RCW\,120 and its ionizing star, and using the \citet{VinDeKLam01} prescription for the mass-loss rate of the star, we find that the position and shape of the inner SWB arc and the outer \ion{H}{ii} region arc are very similar to what is observed, i.e.~the observed inner arc is exactly where we expect it to be if it traces the edge of a SWB.
A model without a stellar wind (simulation HV04) cannot match the observations if dust and gas are dynamically coupled because it does not produce an inner arc of mid-IR emission.

Quantitatively comparing simulation with observation, we predict that the 24 $\mu$m emission should decrease exponentially with increasing distance from the star, whereas the observed decrease is slower for RCW\,120.
We argue that this arises because of either stochastic radiative heating of very small grains that is not included in our model \citep[cf.][]{PavKirWie13}, or a non-axisymmetric geometry of the \ion{H}{ii} region.
We also predict a much sharper outer arc/ring of emission at longer wavelengths than is observed in RCW\,120.
This is because our simulations have a thin and unbroken outer shell, whereas the observations show a much broader and clumpy outer shell (a known limitation of our axisymmetric model).
3D simulations with substructure would agree much better with the observations of the outer shell \citep{WalWhiBisEA15}.
We find that a flatter grain-size distribution with power law exponent $q=2$ matches the data better than the more standard $q\approx3.3-3.5$.

Our results suggest that IR arcs, commonly seen around O stars in \ion{H}{ii} regions, reveal the extent of stellar wind bubbles.
Further work is required to distinguish our displacement model from other explanations of these arcs that do not involve stellar winds, namely the dust-wave model \citep{OchCoxKriEA14} or the mass-loaded-wind model \citep{McKVanLaz84}.
If our model is correct, it opens a new observational window on SWBs, and can provide an additional (and much-needed) constraint on mass-loss rates from O stars by measuring the sizes of SWBs.
Detailed observations of the IR arcs, together with simulations and the post-processing we have done here, could in future constrain the physics of the boundary layer at the contact discontinuity between SWB and \ion{H}{ii} region, in particular if turbulent mixing and/or thermal conduction are active.

\begin{acknowledgements}

JM acknowledges support from the Deutsche Forschungsgemeinschaft priority program 1573, Physics of the Interstellar Medium.
TJ Haworth is funded by the STFC consolidated grant ST/K000985/1.
TJ Haworth is grateful for support from the visitor program of the Deutsche Forschungsgemeinschaft priority program 1573, Physics of the Interstellar Medium.
VVG acknowledges the Russian Science Foundation grant 14-12-01096.
SM gratefully acknowledges the receipt of research funding from the National Research Foundation (NRF) of South Africa.
TJ Harries acknowledges funding from STFC Consolidated Grant ST/M00127X/1.
The authors gratefully acknowledge the computing time granted by the John von Neumann Institute for Computing (NIC) and provided on the supercomputer JUROPA at J\"ulich Supercomputing Centre (JSC).
This research has made use of the NASA/IPAC Infrared Science Archive, which is operated by the Jet Propulsion Laboratory, California Institute of Technology, under contract with the National Aeronautics and Space Administration.
This research has made use of NASA's Astrophysics Data System.
We thank the referee for useful suggestions to improve the discussion of our results.

\end{acknowledgements}

\bibliographystyle{aa}
\bibliography{./refs}

\end{document}